\documentclass[a4paper,11pt]{article}
\pdfoutput=1 

\usepackage{jcappub}
\usepackage[utf8]{inputenc}
\usepackage[T1]{fontenc}

\usepackage{multicol}
\usepackage{placeins}

\usepackage{subcaption}
\usepackage{siunitx}

\newcommand{\HI}{H\,{\sc i}}
\newcommand{\Lya}{Lyman-$\alpha$}

\newcommand{\comment}[1]{}

\usepackage{bm}

\title{Lyman-$\alpha$ polarization from cosmological ionization fronts: II. Implications for intensity mapping}

\author[a,b]{Emily Koivu,}
\author[a,b]{Heyang Long,}
\author[a,c,d]{Yuanyuan Yang,}
\author[a,b,c]{and Christopher M. Hirata}
\affiliation[a]{Center for Cosmology and AstroParticle Physics, The Ohio State University, 191 West Woodruff Avenue, Columbus, OH 43210, USA}
\affiliation[b]{Department of Physics, The Ohio State University, 191 West Woodruff Avenue, Columbus, OH 43210, USA}
\affiliation[c]{Department of Astronomy, The Ohio State University, 140 West 18th Avenue, Columbus, OH 43210, USA}
\affiliation[d]{Khoury College of Computer Science, Northeastern University, 440 Huntington Ave, Boston, MA 02115, USA}
\emailAdd{koivu.1@osu.edu}

\abstract{This is the second paper in a series whose aim is to predict the power spectrum of intensity and polarized intensity from cosmic reionization fronts. After  building the analytic models for intensity and polarized intensity calculations in paper I, here we apply these models to simulations of reionization. We construct a geometric model for identifying front boundaries, calculate the intensity and polarized intensity for each front, and compute a power spectrum of these results. This method was applied to different simulation sizes and resolutions, so we ensure that our results are convergent.
We find that the power spectrum of fluctuations at $z=8$ in a bin of width $\Delta z=0.5$ ($\lambda/\Delta\lambda=18$) is $\Delta_\ell \equiv [\ell(\ell+1)C_\ell/2\pi]^{1/2}$ is $3.2\times 10^{-11}$ erg s$^{-1}$ cm$^{-2}$ sr$^{-1}$ for the intensity $I$, $7.6\times10^{-13}$ erg s$^{-1}$ cm$^{-2}$ sr$^{-1}$ for the $E$-mode polarization, and  $5.8\times10^{-13}$ erg s$^{-1}$ cm$^{-2}$ sr$^{-1}$ for the $B$-mode polarization at $\ell=1.5\times10^4$. 
After computing the power spectrum, we compare results to detectable scales and discuss implications for observing this signal based on a proposed experiment. We find that, while fundamental physics does not exclude this kind of mapping from being attainable, an experiment would need to be highly ambitious and require significant advances to make mapping \Lya\ polarization from cosmic reionization fronts a feasible goal.
}


\date{\today}

\begin{document}
\maketitle
\flushbottom

\section{Introduction}

Lyman-$\alpha$ line intensity mapping is expected to be a probe for the epoch of reionization -- the period of cosmic history when the primordial stars and galaxies reionized the neutral hydrogen in the intergalactic medium \cite{2013ApJ...763..132S,Pullen_2014,2017ApJ...848...52H,2022arXiv221009612S}. In an earlier paper \cite{PaperI} (hereafter Paper I), we created a model of polarized \Lya\ emission from a plane-parallel cosmic ionization front, including a treatment of polarized intensity emitted from the fronts. This model incorporates a rigorous treatment of the ionization structure, the thermal structure with a multi-temperature plasma, and the directional dependence of scattering cross sections for polarized photons.
In this paper, we use this model to create an intensity map for the period of reionization on a realistic cosmological history (see \cite{https://doi.org/10.48550/arxiv.1709.09066} for an overview of intensity mapping; for a range of redshifts and techniques for \Lya\ line intensity mapping experiments see HETDEX \cite{2008ASPC..399..115H,2021ApJ...923..217G}, PAU \cite{2021MNRAS.501.3883R}, and SPHEREx \cite{2014arXiv1412.4872D}). From this intensity calculation, we aim to compute the power spectrum of total and polarized \Lya\ emission from the ionization fronts.
The power spectrum of \Lya\ polarization from the ionization fronts is expected to follow the distribution of scales of the reionization bubbles. By tracing the evolution of this intensity signal, one can learn about the geometric evolution of sources in the reionization period. 

A second goal of this paper is to do a first assessment of the detectability of the polarized Lyman-$\alpha$ signal from ionization fronts, comparing it to the previously studied signal from galaxies \cite{Loeb_1999,Dijkstra_2008}, as well as to the sensitivity of plausible surveys with near-term technology.
For both purposes, we compare the results of our calculations to the analysis of Mas-Ribas \& Chang \cite{MasRibasChang}, which is the most extensive prior investigation of Lyman-$\alpha$ polarized intensity mapping from the epoch of reionization. In their work, Mas-Ribas \& Chang discuss polarized \Lya\ emission detectability relative to a proposed sensitivity scale based on current instrumentation capacity, though they consider radiation from galaxy haloes and not ionization fronts themselves. By comparing to these results to our own findings, we will be able to assess detectability of these cosmic reionization fronts relative to emission from galaxies, and discuss the sensitivity scales a future polarized intensity mapping survey may need to explore this interesting time in our universe's history. 

This paper is organized as follows: Section \ref{sec:Method} describes the process of generating reionization simulations, extracting and characterizing ionization fronts, calculating the intensity and polarized intensity of each front, and creating a power spectrum for the intensities from the fronts. This section will also contain convergence tests as a function of simulation box size and resolution. Section \ref{Sec:results} shows results of the power spectra for the intensity and polarized intensity, and compare these results to the results of Mas-Ribas \& Chang. Finally, we discuss our findings and future areas of interest in polarized intensity mapping in Section \ref{sec:Discussion}. In this paper, we will continue using the notation from Paper I.

\section{Methodology}

In this section we describe the process of generating a simulation of reionization ionization using {\sc 21cmFAST} \cite{Murray2020,10.1111/j.1365-2966.2010.17731.x}, constructing a model of the ionization fronts, characterizing these fronts, and creating the intensity and polarized intensity power spectrum. For the remainder of this paper, we adopt the \textit{Planck 2018} ``TT,TE,EE+lowE+lensing'' $\Lambda$CDM cosmological parameters \cite{2020A&A...641A...6P}; this is consistent with assumptions made in Paper I.

\label{sec:Method}

\subsection{{\sc 21cmFAST} Simulation}
\label{ss:21cmFAST}

\begin{figure}
    \centering
    \includegraphics[width=.7\textwidth]{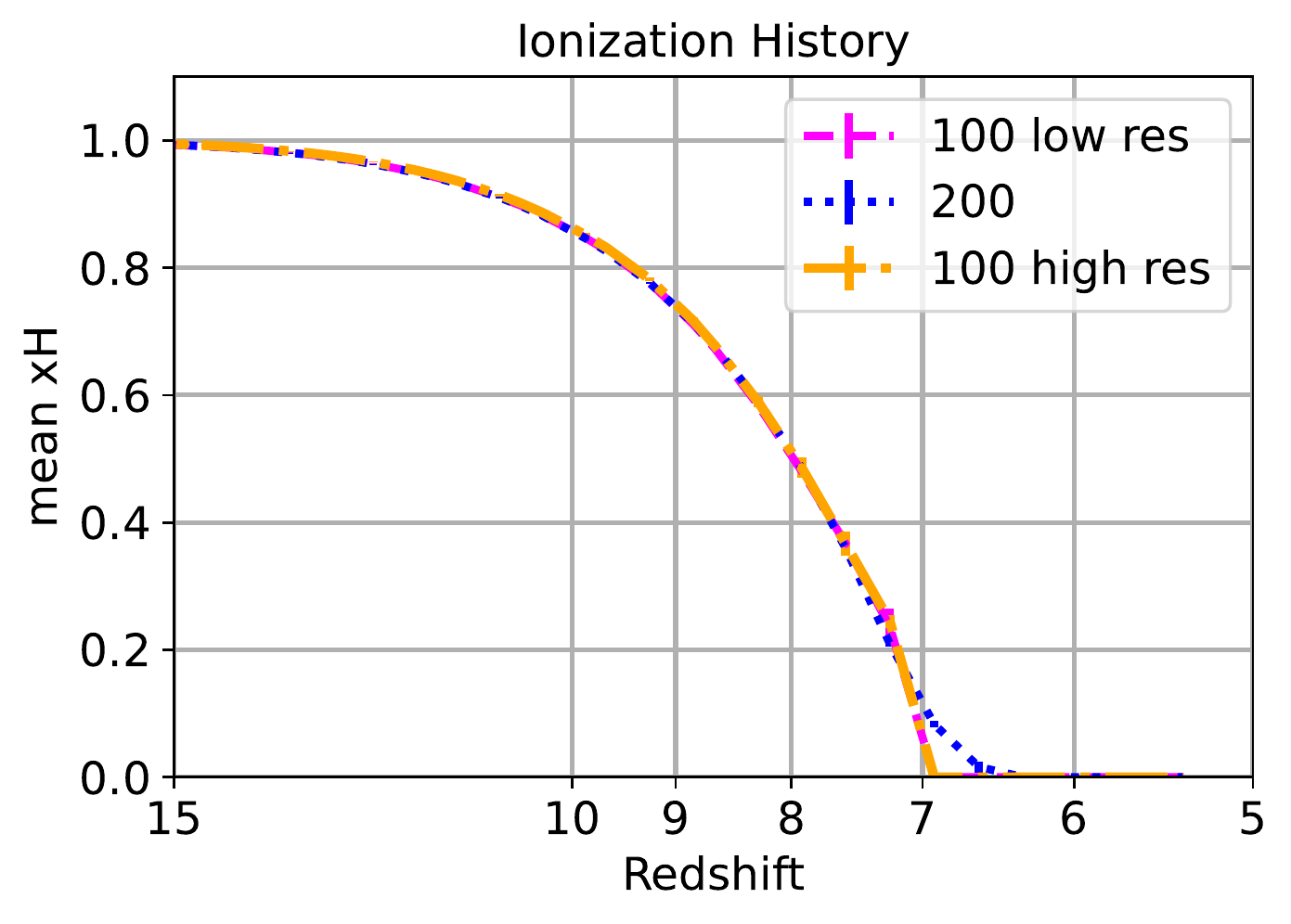}
    \caption{\large The reionization history (neutral fraction versus redshift) for our simulation boxes. The simulations run from $z=35$ to $z=5$, but are almost fully neutral prior to $z=15$ so the early history is omitted from this plot. All boxes display half ionization at $z=7.91$. }
    \label{fig:ReionHist}
\end{figure}

In order to achieve a realistic cosmological  to observe ionization from the reionization era, we used the {\sc 21cmFAST} semi-numerical code to produce 3-dimensional coeval cubes. These cubes contain a vast array of information, including information about the hydrogen density, the ionization fraction, and the metagalactic
hydrogen ionization rate $\Gamma_{12}$ (the photoionization rate per H{\sc\,i} atom, in units of $10^{-12}$ s$^{-1}$) values of each cell which will be of interest for our calculations. This code uses initial conditions to create a perturbed field at $z= 35$. From this initially perturbed field and a set of desired astrophysical processes and constraints, this simulation produces a sequence of ionized boxes as a function of redshift. Using this information, {\sc 21cmFAST} will continue to create perturbed fields and ionization boxes iteratively by evolving forward in time, until reaching the desired redshift for which we wish to conduct our calculations. For our purpose, we wish to investigate $z=8$ though we simulation ionization for redshifts up to $z=5$ (i.e., after the end of reionization). Our interest in $z =8$ is because  reionization is at a peak around $z=8$ so the \Lya\ polarization signal from fronts at this time is representative and might be the strongest \cite{2017MNRAS.465.4838G,2018ApJ...864..142D}, which also reflects the model produced in Paper I. Figure \ref{fig:ReionHist} shows the ionization history for simulations computed. All simulations show a similar reionization history. 

The {\sc 21cmFAST} simulations are created using an efficiency factor of $\zeta = 45.0$ (defined by eq. 2 of Ref. \cite{Greig_2015}). 
Using this efficiency factor allowed for the midpoint redshift of reionization to agree with that from best-fit Planck cosmology \cite{2020A&A...641A...6P} within 1$\sigma$, while still not requiring unreasonably high efficiency. We also allow inhomogeneous recombinations, and intergalactic medium spin temperature fluctuations (although the latter do not affect the ionization structure and hence they have no impact on the results of this paper). 

While investigating the analysis pipeline, these simulated coeval cubes produce a (100 Mpc)$^3$ volume with each cell having a (1 Mpc)$^3$ volume. In verifying the validity of our results, we also implement a (100 Mpc)$^3$ cube with cells of (0.5 Mpc)$^3$ to compare intensity results for simulations with increased resolution, and (200 Mpc)$^3$ cube with (1 Mpc)$^3$ volume cells to compare results on a larger data set. 

\subsection{Ionization Fronts Shape Extraction}

\begin{figure}
\centering
\includegraphics[width=\textwidth]{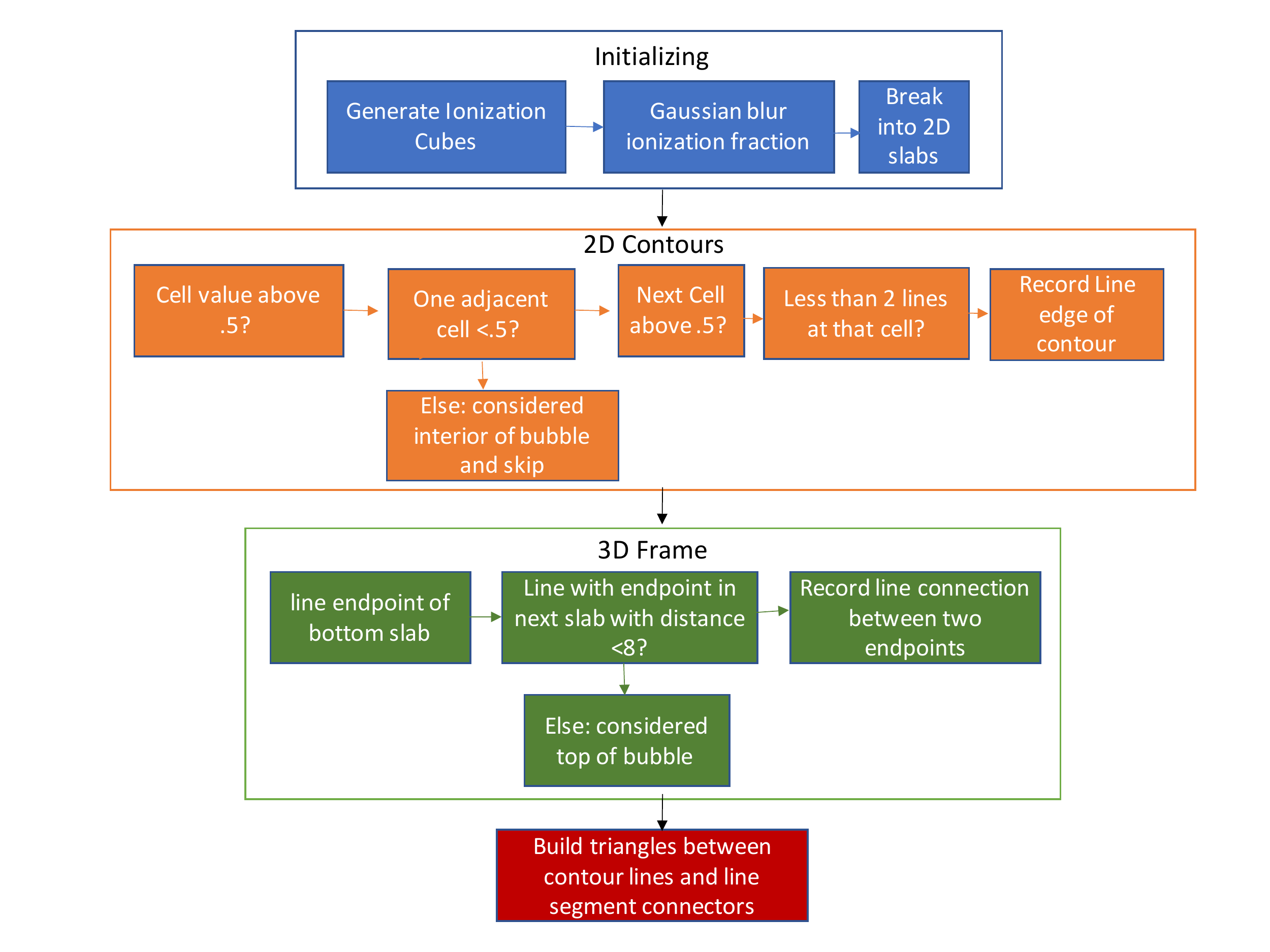}
\caption{\large Flowchart of the ionization front modeling method.
\label{fig:FC}}
\centering
\end{figure}

\begin{figure}
    \centering
    \includegraphics{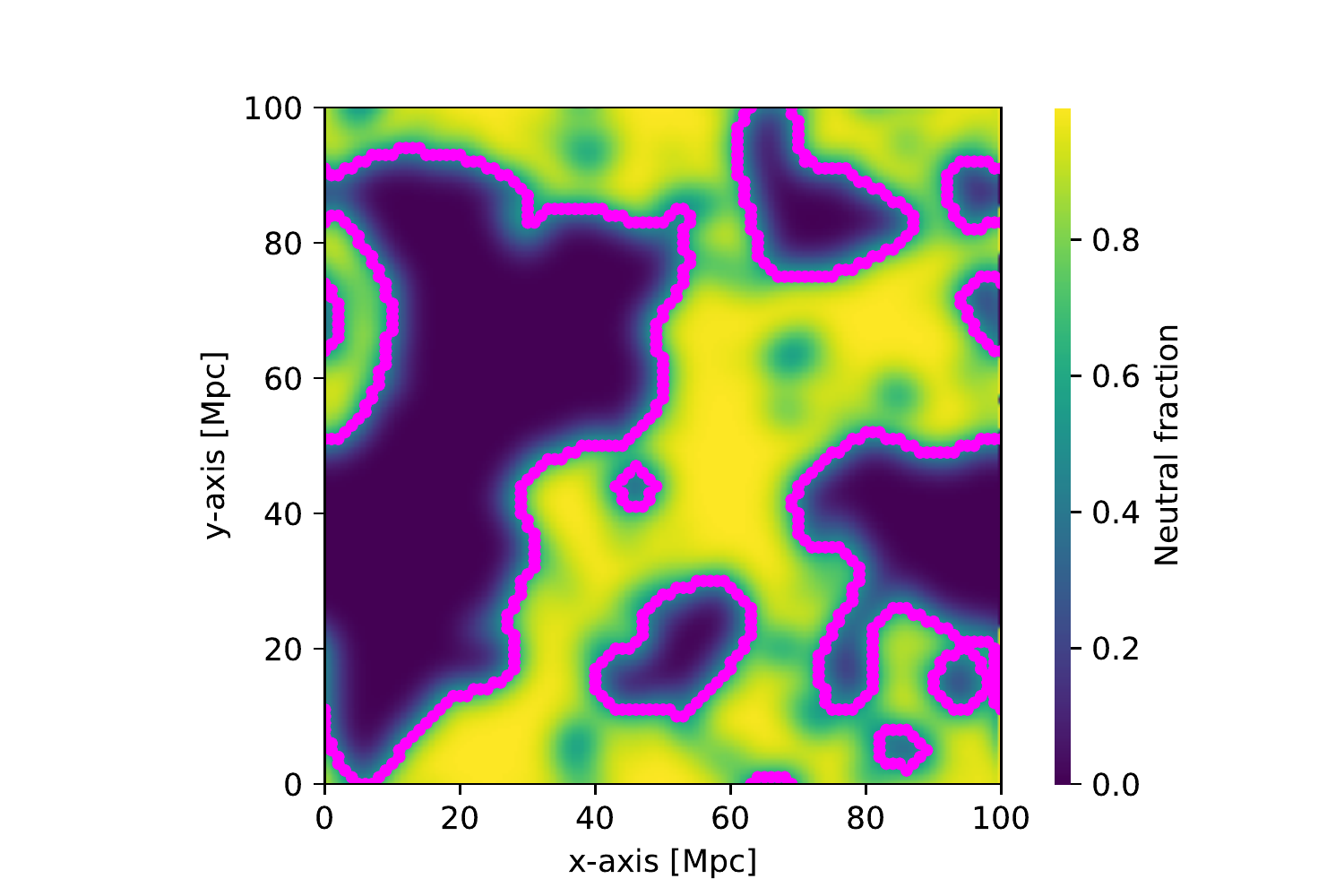}
    \caption{\large 2D Contour Example. Background represents the ionized hydrogen fraction for one slab of the simulated box, with the gaussian blur applied. The magenta curves overlaid show the recorded contours for the ionization fronts.}
    \label{fig:2DContour}
\end{figure}

In order to characterize the ionization fronts from {\sc 21cmFAST} simulations, we need a method to locate the fronts. A summary of this method is found in Figure \ref{fig:FC}, and is detailed below. In the three dimensional ionization cube for each redshift, each output cell in the cube is given an ionization fraction. We first use a Gaussian convolution to smooth over the simulation cube. This gives us the most prominent features of the ionization bubbles while ignoring any jagged edges or small bubbles at the resolution scale that may not be properly represented. (We perform a resolution convergence test in Section~\ref{ss:conv}.) This Gaussian blur is set to smooth over $\sigma = 2.5$ cells with periodic boundary conditions. An example of the blurred ionization fraction is shown in the background of Figure \ref{fig:2DContour}.

The next step in finding the fronts was to break up the cubes into 2D slabs parallel to the $xy$-plane and determine the contours of ionization fronts for each slab.\footnote{We chose slabs in the plane of the sky as seen from the observer -- who looks along the $z$-axis -- so that the numerical procedure does not break the 4-fold discrete rotational symmetry around the line of sight. This symmetry protects the overall mean of the linear polarization from getting a contribution: $\langle Q\rangle = \langle U \rangle$ = 0.} An ionization front is defined as a cell with ionization fraction above 0.5 and at least one nearest neighbor with ionization fraction less than 0.5. Upon finding a potential point on the ionization front, we investigate if there is an adjacent point (including diagonal neighbors) that also meets our ionization front criteria, and record the pair of points that make a line segment in the ionization front contour. For each ionization front cell, we limit the number of lines to two. The two dimensional contour for the example blurred ionization fraction is shown in the magenta curves for Figure \ref{fig:2DContour}.

After finding the two dimensional ionization front contours in each slab, we take each point on the contour of a slab and search for the nearest ionization front point on the next slab up. If the distance to this nearest point is less than 8 cells (our threshold for being in the same ionization bubble), then we record the points in each slab as a line connecting the two contours. If a contour has no point within this threshold distance in the next slab, we determine this to be the top region of the ionization bubble. We do this procedure for all contours segments in all slabs. 

\begin{figure}
\centering
    \subfloat[]{
    \includegraphics[scale=.48]{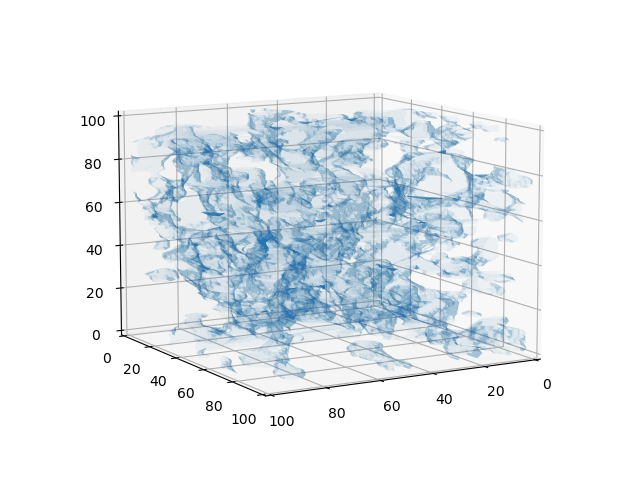}
    }
    \subfloat[]{
    \includegraphics[scale=.48]{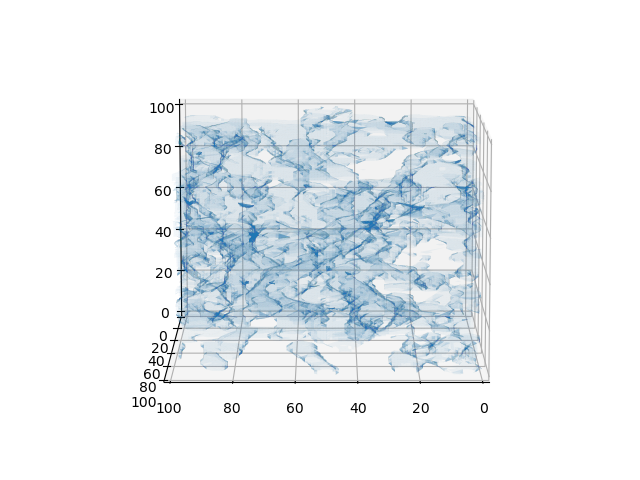}
    }
    \hspace{0mm}
    \subfloat[]{
    \includegraphics[scale=.48]{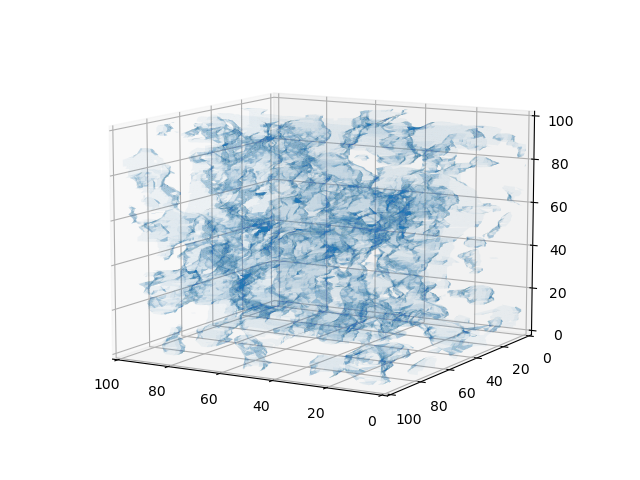}
    }
    \hspace{0mm}
    \caption{\large Example of triangles that form the ionization boundaries plotted from three different viewing angles: 60$^\circ$, 90$^\circ$, and 120$^\circ$ from the $x$-axis in the $xy$-plane. All are taken with a 15$^\circ$ incline in the x-z plane to give a slight top down view of the surfaces. The triangles each have 30\% opacity to allow for investigation of inner regions; therefore, more opaque regions have more triangles and, by extension, ionization front boundaries along this line of sight. This box has a volume of (100 Mpc)$^{3}$ with a cell size of (1 Mpc)$^{3}$ taken at redshift $z=8$.}
    \label{fig:3d Fronts 3 Angles}
\end{figure}

The final step in determining the full three dimensional model of the ionization fronts is to construct triangles that map out the surface. An example plot of these surfaces from three different viewing angles is shown in Figure \ref{fig:3d Fronts 3 Angles}. To find the triangles that best match our surface, we take each contour segment in a slab and determine the lines for each endpoint that mapped to the contour in the next slab up. We then determined all the contour segments that were needed to connect from the endpoints in the upper contour. Using all of these line segments, from top and bottom slabs, as well as the lines connecting them, we constructed triangles that build up the surface of the enclosed area. For each triangle, we calculated the area $A_\Delta$, the centroid ${\bf r}_{\rm centroid}$, and normal vector $\hat{\bf n}$ to determine the outward direction of the front. A triangle has two possible normal directions $\pm\hat{\bf n}$; we choose the direction pointed toward the neutral region, which we determine by assuming one $\hat{\bf n}$ to be true and assessing the ionization fraction for the cell closest to $\sqrt{3}$ cells from the center of the front triangle in $\hat{\bf n}$ direction. If the ionization fraction in this cell is below 0.5, we proceed with this direction as $\hat{\bf n}$; otherwise, we assign the outward direction as $-\hat{\bf n}$. In this way, we are assigning the direction as towards the neutral region and away from the ionized region.

\begin{figure}
    \centering
    \subfloat[z=35]{\includegraphics[width=50mm]{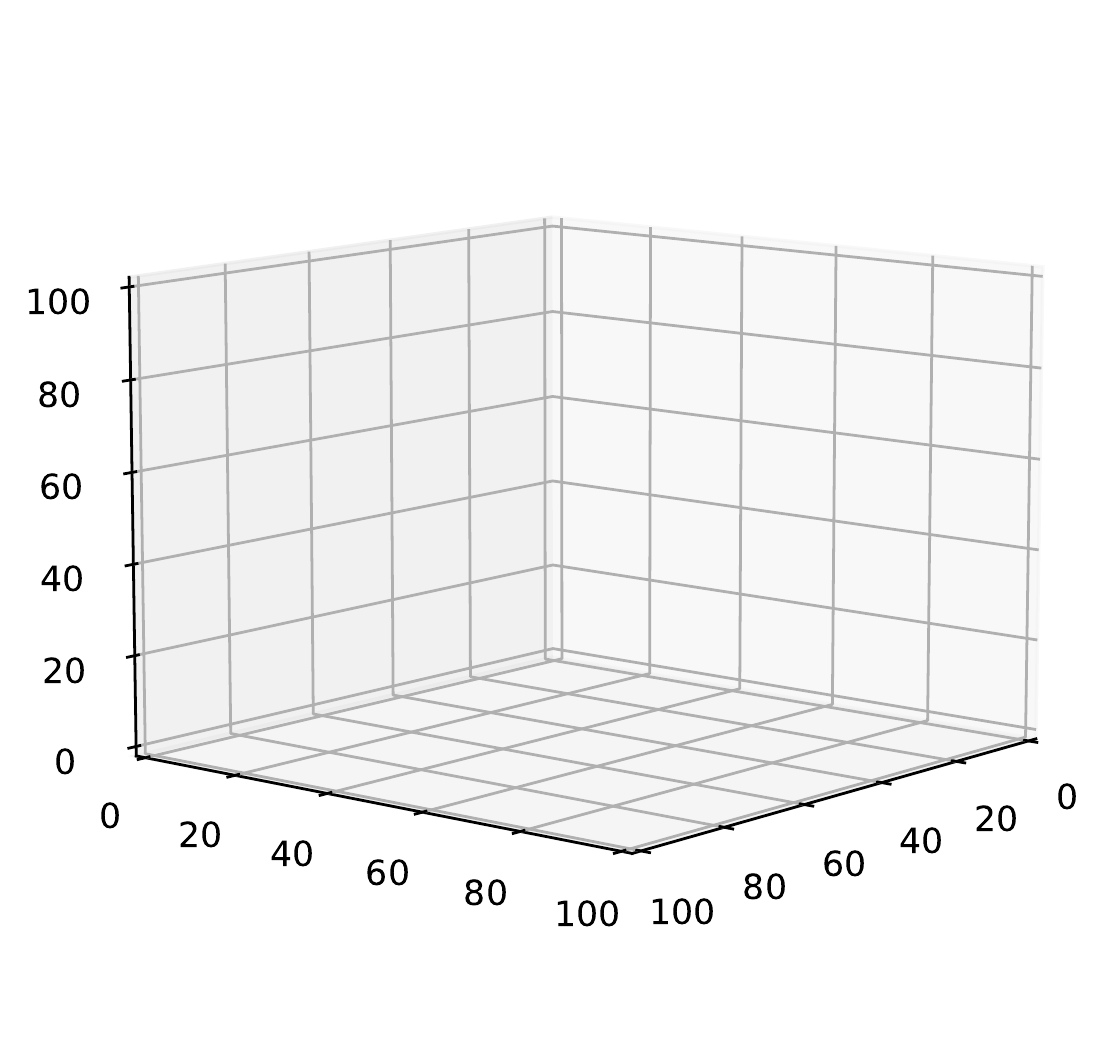}}
    \subfloat[z=13.5]{\includegraphics[width=50mm]{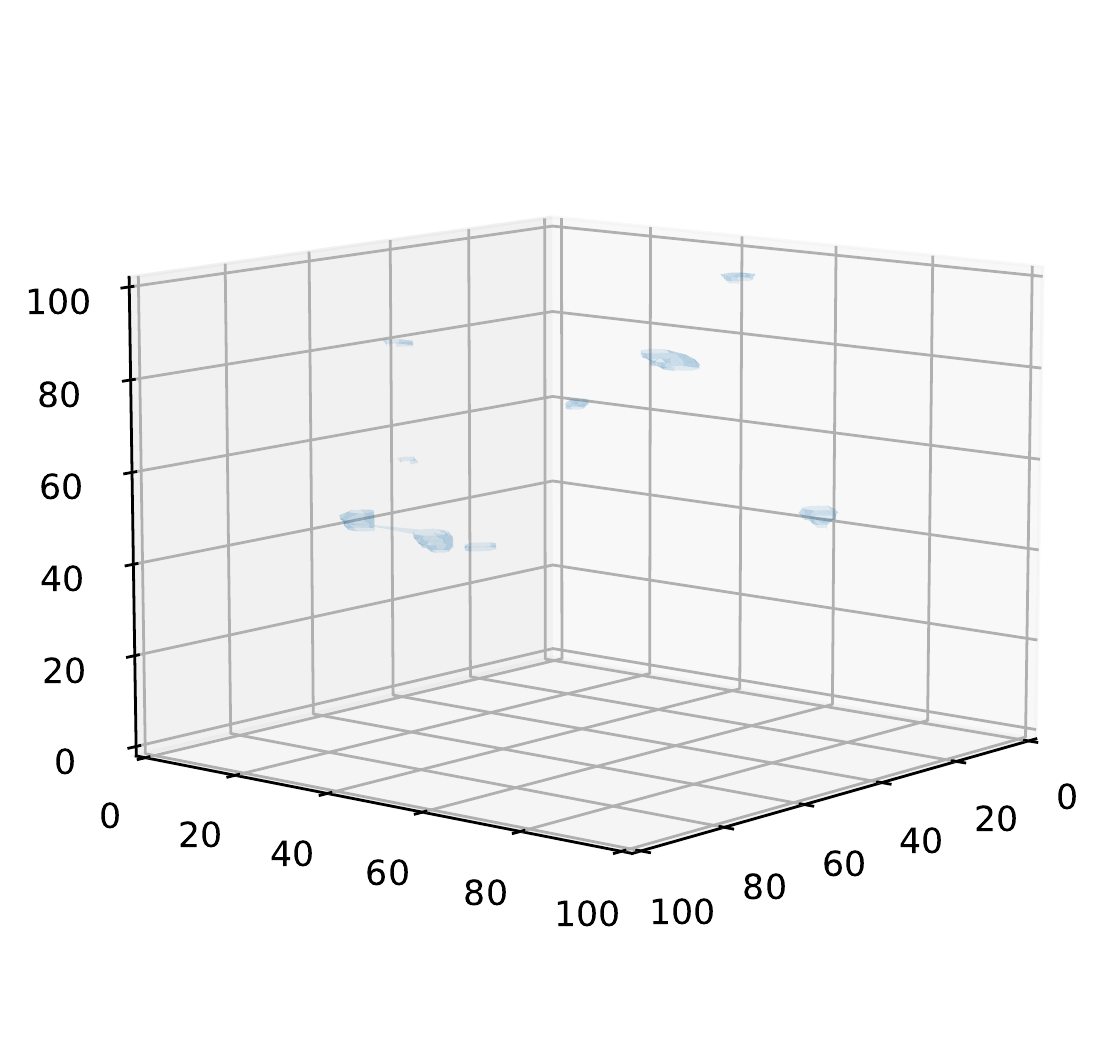}}
    \subfloat[z=12.5]{\includegraphics[width=50mm]{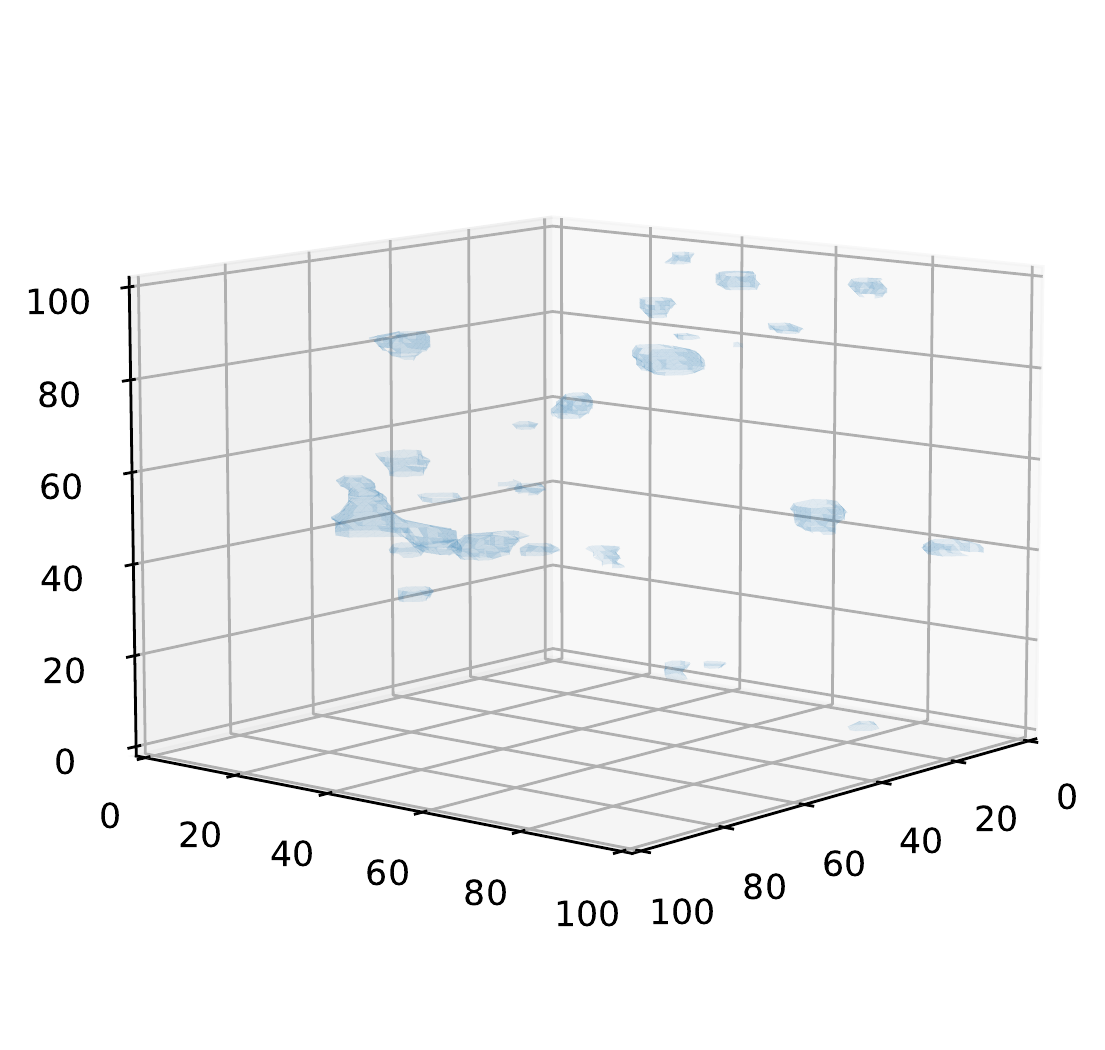}}
    \hspace{0mm}
    \subfloat[z=12]{\includegraphics[width=50mm]{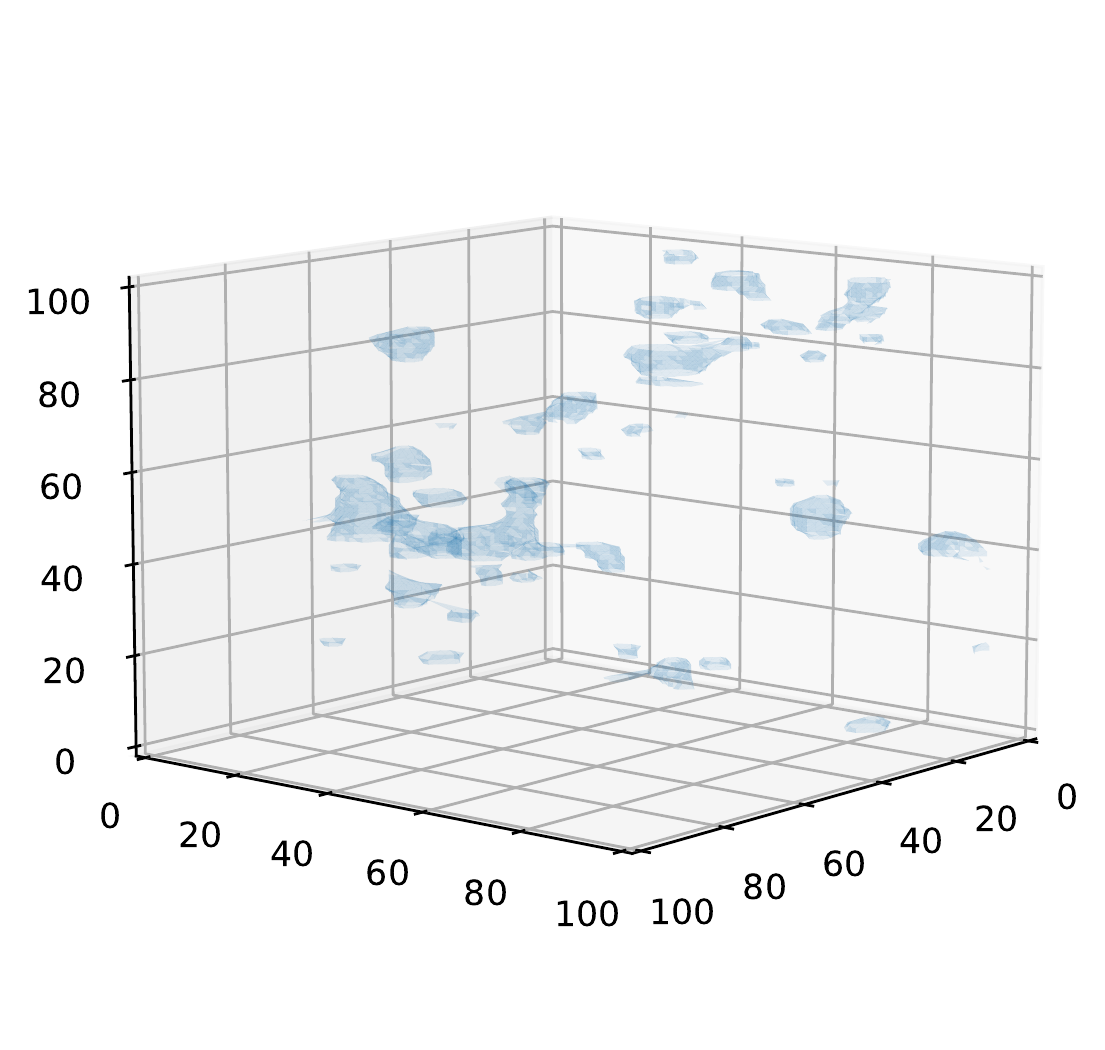}}
    \subfloat[z=11]{\includegraphics[width=50mm]{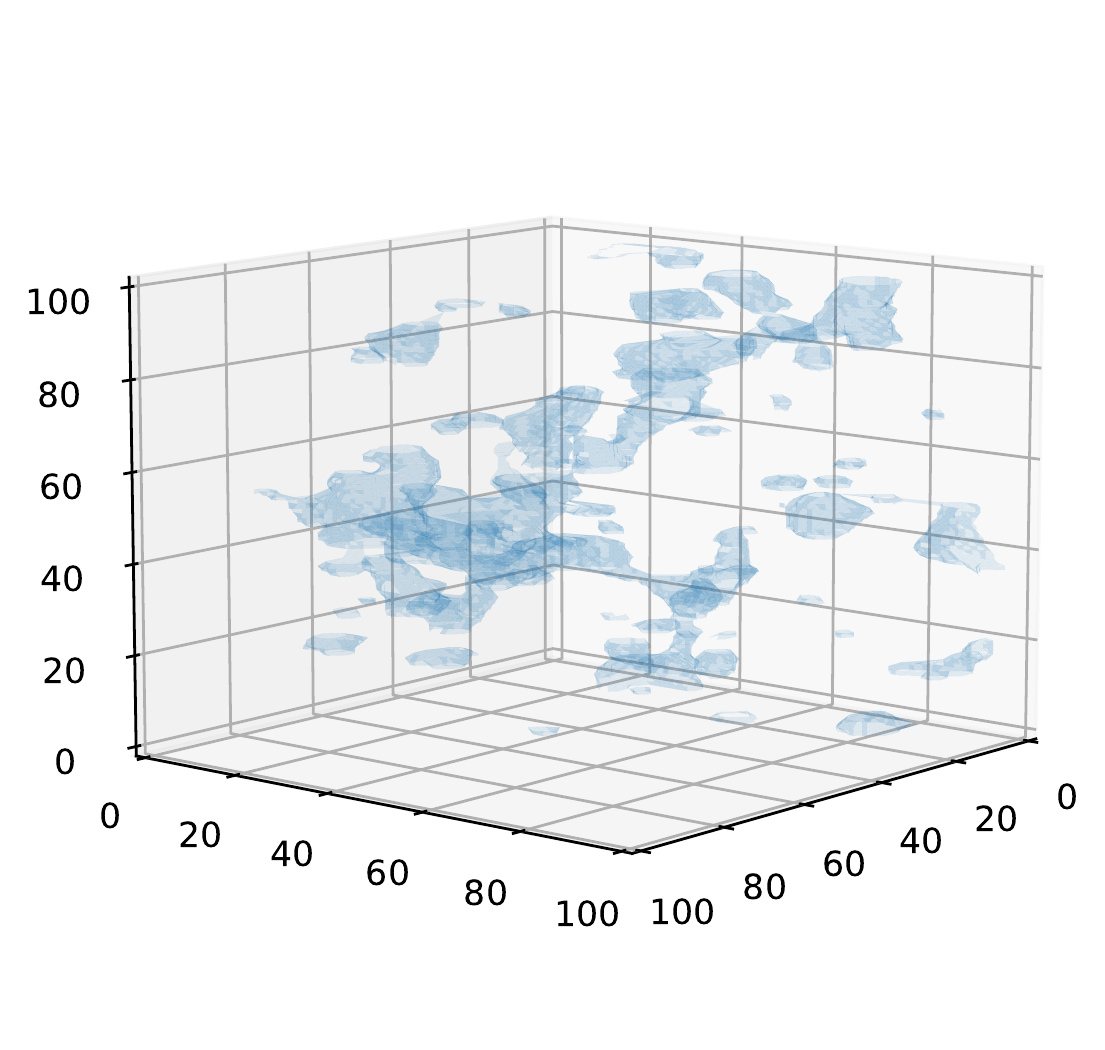}}
    \subfloat[z=10]{\includegraphics[width=50mm]{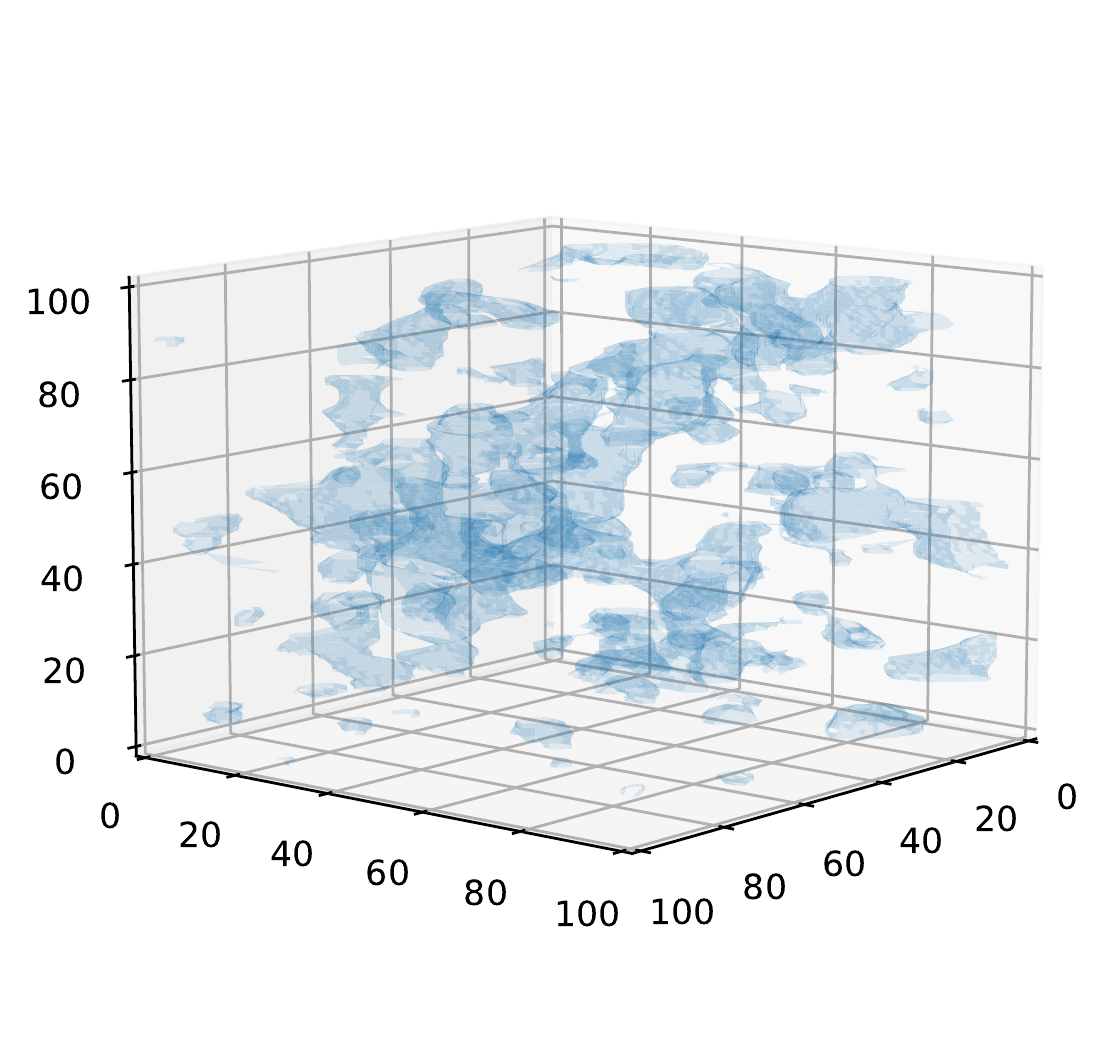}}
    \hspace{0mm}
    \subfloat[z=9]{\includegraphics[width=50mm]{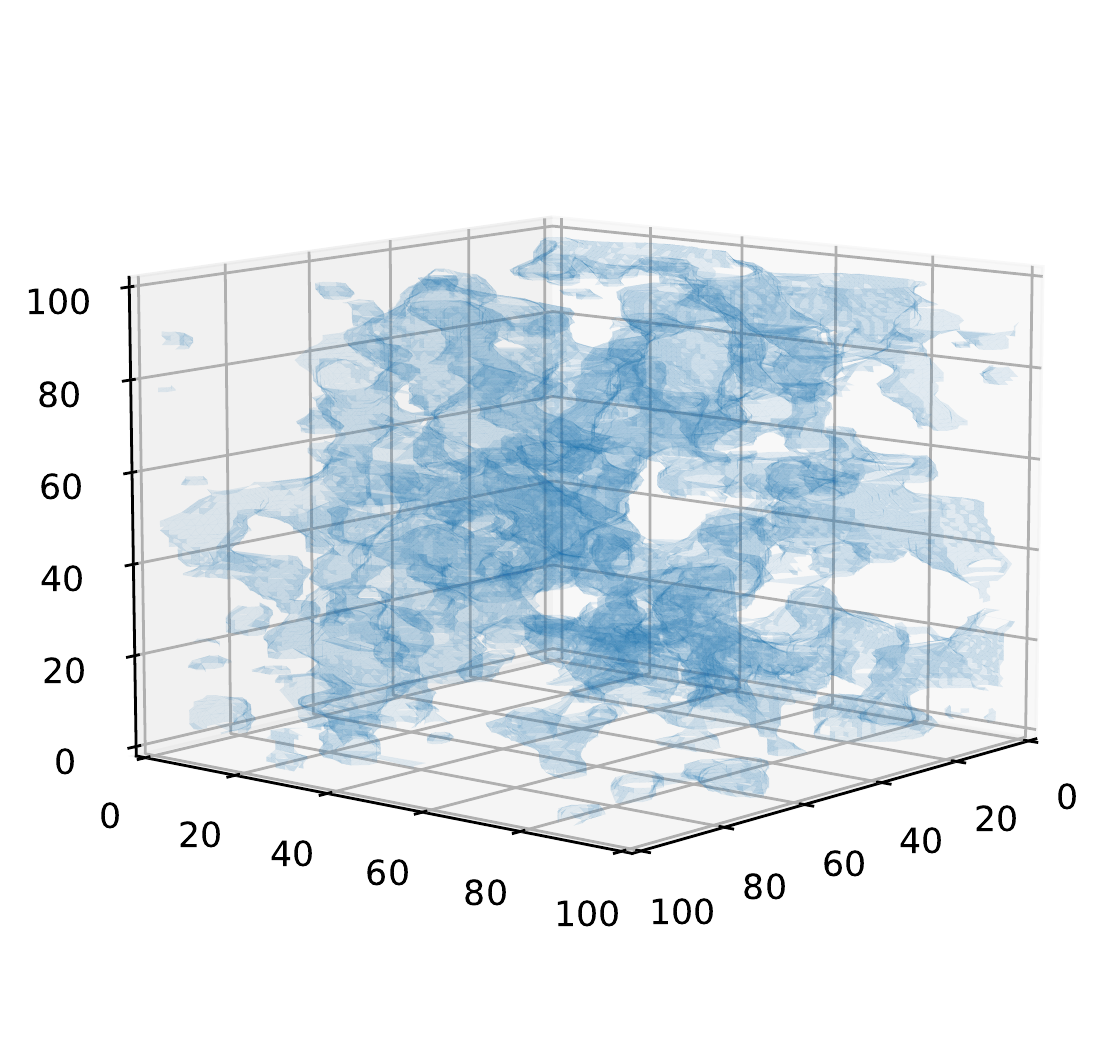}}
    \subfloat[z=8]{\includegraphics[width=50mm]{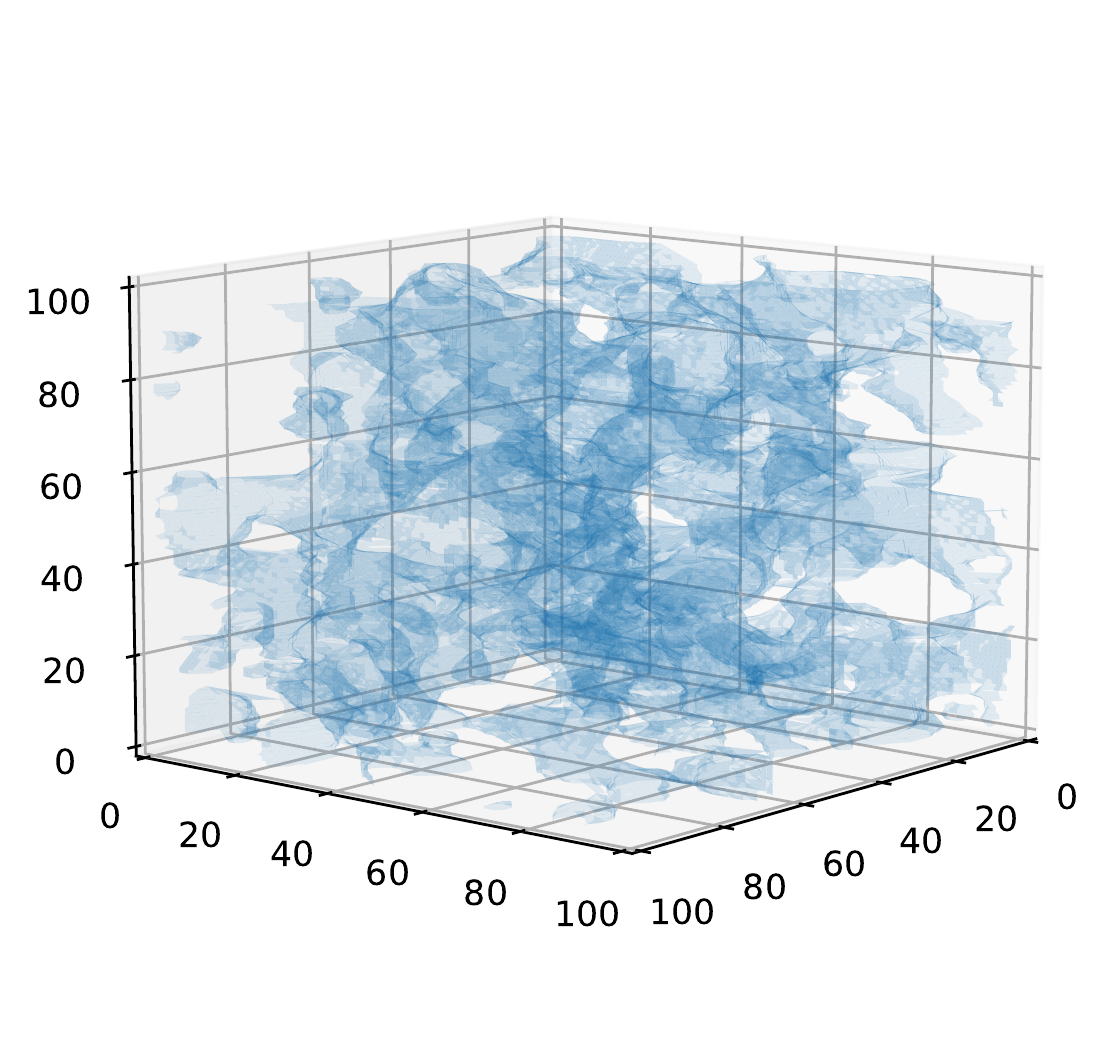}}
    \subfloat[z=7]{\includegraphics[width=50mm]{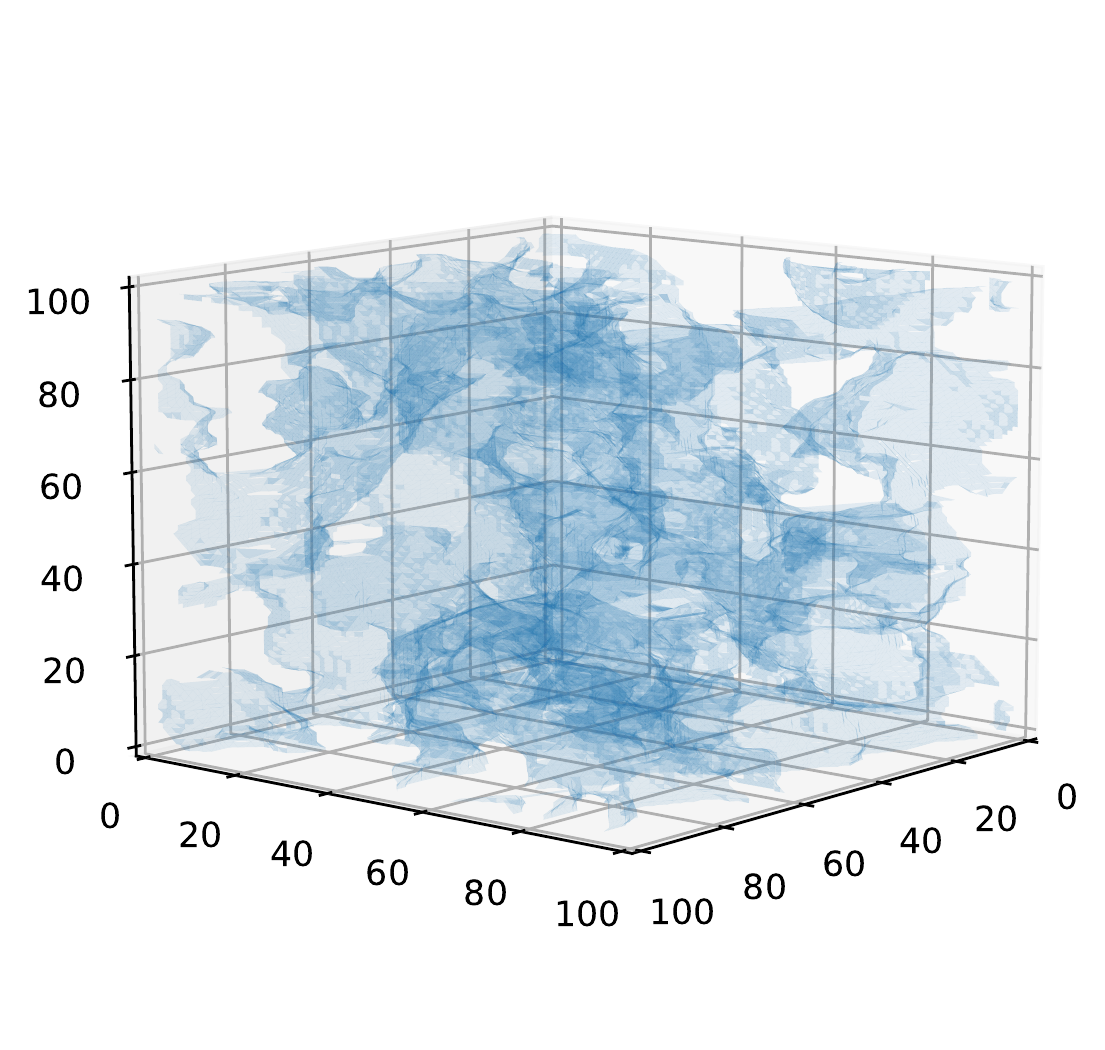}}
    \caption{Time Evolution of ionization fronts. Shown are ionization fronts from a single simulation of a (100 Mpc)$^3$ box with cell size of 1 Mpc$^3$. The figures demonstrate decreasing redshifts, starting at redshift 35, then proceeding to 13.5, 12.5, 12, 11, 10, 9, 8, and ending with figure (i) at redshift 7.}
    \label{fig:time evolve ionization fronts}
\end{figure}

As stated in Section \ref{ss:21cmFAST}, the {\sc 21cmFAST} simulation tool evolves perturbations forward in time from redshift $z=35$ to the desired redshift. To verify that the simulation and front modeling tools are properly functioning, a forward evolution of the three dimensional triangle fronts is shown in Figure~\ref{fig:time evolve ionization fronts}. This figure demonstrates the growth of ionization bubbles in our cosmological simulation as time progresses, leading to an intricate structure of ionization fronts in the final redshift cube. 

\subsection{Ionization front characteristics}
\label{sec:IFC}

In order to make use of the results from Paper I, we want to extract the blackbody temperature, front velocity and neutral hydrogen density for each triangle on the ionization front.

The blackbody temperature of the incident radiation is treated as fixed and not computed on a cell-by-cell basis. We simulate the spectral energy distribution by star-forming galaxies using {\sc Starburst99} \cite{1999ApJS..123....3L} and tune the star-forming galaxy model parameters such that the number of ionizing photons per baryon matches that in the setup of {\sc 21cmFAST}. Then we fit the emission energy spectrum as emitted by a black body with $T_{\rm bb}=4.78\times 10^4$\,K, corresponding to a stellar population with constant star formation rate, metallicity $0.008\,Z_\odot$, and an initial mass function given by Ref. \cite{2001MNRAS.322..231K}:
\begin{equation}
    \xi(m)= \begin{cases}
    m^{-1.3}, & 0.1\,M_{\odot} < m < 0.5\,M_{\odot} \\
    m^{-2.3}, & 0.5\,M_{\odot} < m < 107\,M_{\odot}.
    \end{cases}
\end{equation}
This treatment neglects the hardening of the radiation as it is filtered through the ionized intergalactic medium (which has a nonzero column density of H{\sc\,i} and He{\sc\,i}) and the contribution of ionizing photons from intergalactic recombinations (see, e.g., Ref.~\cite{1996ApJ...461...20H} for a discussion). However, we expect it to be reasonable for a first calculation of the Lyman-$\alpha$ emission from the fronts.

Second, we want the density of the gas (parameterized by $n_{\rm H}$) into which the front is propagating. Since {\sc 21cmFAST} does not provide the number density directly, but it does store the matter overdensity data, we calculate the hydrogen number density by
\begin{equation}
    n_{\rm H} = {\bar{n}_{\rm H}} (1+\delta_{\rm b}) \approx \frac{\Omega_{\rm b,0}X_{\rm H}\rho_{\rm crit,0}(1+z)^3}{m_{\rm H}} (1+\delta_{\rm m}),
\end{equation}
where $\Omega_{\rm b,0}$ is the baryon abundance today; $X_{\rm H} = 0.76$ is the hydrogen mass fraction; $\rho_{\rm crit,0} = 8.53\times10^{-30} \ \rm g \hspace{1mm} cm^{-3}$ is the critical density today for $H_0 = 67.36$ km/s/Mpc, $m_{\rm H}$ is the hydrogen atom mass, and $\delta_{\rm m}$ is the matter overdensity we could extract from simulation box. We neglect the difference between $\delta_{\rm b}$ and $\delta_{\rm m}$ since we are on scales large compared to the Jeans length.

Finally, we calculate the front velocity $U$ by considering the photoionization of \HI\ by ionizing flux at the boundary of ionizing bubbles, which satisfies the following equation:
\begin{equation}
    \frac{n_{\rm H}(1+f_{\rm He})U}{1-U/c} = \int_{I_{\rm H}/h}^{\infty} d\nu\frac{F_{\nu}^{\rm inc}}{h\nu} = \frac{\Gamma_{\rm HI}{\rm (ionized \,side)}}{\bar{\sigma}_{\rm HI}},
\end{equation}
where $f_{\rm He} = (1-X_{\rm H})/4X_{\rm H} = 0.079$ is the helium-to-hydrogen number ratio, $c$ is the speed of light, $\Gamma_{\rm HI} = 10^{-12}\Gamma_{12}\,{\rm s}^{-1}$ is the photoionization rate, and $\bar{\sigma}_{\rm HI}$ is the H{\sc \,i} photoionization cross section averaged over the incident spectrum. This can be expressed as
\begin{equation}
    \bar{\sigma}_{\rm HI} = \frac{\int_{I_{\rm H}/h}^{4I_{\rm H}/h} d\nu F_{\nu}^{\rm inc} \sigma_{\rm HI}(\nu)/h\nu}{\int_{I_{\rm H}/h}^{4I_{\rm H}/h} d\nu F_{\nu}^{\rm inc}/h\nu},
\end{equation}
where $I_{\rm H} = 13.6$ eV is the ionization energy of hydrogen, $h$ is Planck's constant, and the incident flux is a rescaled blackbody spectrum, $F_\nu^{\rm inc}\propto B_\nu(T_{\rm bb})$. We use the hydrogenic cross section (e.g., Eq.~2.4 of Ref.~\cite{2006agna.book.....O}).

As it takes about 2 to 200 hours to simulated 100,000 photons depending on the blackbody temperature $T_{\rm bb}$, front velocity $U$, and neutral hydrogen density $n_{\rm H}$, it would be very expensive to re-run the microphysics simulation for every triangle in every ionization front in the whole simulation box. However, $T_{\rm bb}$ of {\sc 21cmFAST} is fixed, and the simulation results in Paper I only depended on $U$ and $n_{\rm H}$. Thus, we can build a 2D interpolation table for intensity and polarized intensity as a function of $U$ and $n_{\rm H}$, and only call the interpolating function for each triangle.

For our interpolation, we fixed $T_{\rm bb}$ and ran set 12 sets of $U$ varied from $7 \times 10^6$ cm/s to $2.7 \times 10^{10}$ cm/s, and 19 sets of $n_{\rm H}$ varied from $\rm 10^{-9} \ cm^{-3}$ to 1 $\rm cm^{-3}$. This corresponds to computation of a total of $12\times 19 = 228$ ionization front models. We use bilinear interpolation from the four nearest points (e.g., Eq.~25.2.66 of Ref.~\cite{1972hmfw.book.....A}). In practice, we interpolate the Legendre polynomial and associated Legendre polynomial coefficients for the probability distribution P($\mu$) and polarization-weighted probability distribution $(Q/I){\rm P}(\mu)$, respectively, as computed using the procedure in Paper I.
After interpolating these coefficients, we can calculate ${\rm P}(\mu)$ and $(Q/I){\rm P}(\mu)$ for each set of $U$, $n_{\rm H}$ and $\mu$. By connecting to the interpolated Lyman-$\alpha$ photon emission rate $n$, we can calculated the intensity and polarized intensity for each given value of $U$, $n_{\rm H}$, and $\mu$.

\subsection{Power spectra of Lyman-$\alpha$ emission}

Our final step is to convert the intensities from each triangle into an overall power spectrum. This is done in two steps: first, we interpolate the triangles onto a rectangular grid; and then we transform to Fourier space to compute the power spectrum.

For the first step, we split each front triangle into $N^2$ discrete sub-triangles (where the integer $N$ depends on the size of the triangle). Each sub-triangle is then placed into a grid cell. This is all straightforward in principle, but requires some book-keeping to be done correctly with 2D triangles and a 3D rectangular grid and to rotate the polarization directions to a common coordinate system; we give the formulae in Appendix~\ref{appendix:split}.
Implementing this operation on all front triangles we obtain the boxes representing the \Lya\ specific intensity $I_{\nu}$ and polarized specific intensity $Q_{\nu}$ and $U_{\nu}$ from ionizing fronts in real space.

Then we do a 3-dimensional Fourier transform for specific intensity in real space $I_{\nu}(n_{r_1},n_{r_2},n_{r_3})$ (N grids in each dimension) to obtain the Fourier space intensity $\tilde I_{\nu}(k_1,k_2,k_3)$ 
\begin{equation}
    \tilde I_{\nu}(k_1,k_2,k_3) = V_{\rm grid}\sum_{r_1,r_2,r_3}I_{\nu}(r_1,r_2,r_3)e^{\frac{2\pi i}{N}(k_1n_{r_1}+k_2n_{r_2}+k_3n_{r_3})}
\end{equation}
 where $V_{\rm grid}$ is the unit grid volume, $n_{r_1},n_{r_2},n_{r_3}$ represent the grid index in each dimension. The power spectrum $P_{I_{\nu}} (\textbf{k})$ is
\begin{equation}
    P_{I_{\nu}} (\textbf{k}) = \frac{|I_{\nu}({\bf k})|^2}{V_{\rm box}}
\end{equation}
where $V_{\rm box}$ is the total proper volume for the simulation box. We use the same normalization in the calculation of polarization quantities $\tilde Q_{\nu}(k_1,k_2,k_3)$ and $\tilde U_{\nu}(k_1,k_2,k_3)$.
In the flat-sky approximation ($l\gg 1$), the $E$ and $B$ modes can be written as a rotation of the $Q$ and $U$ Stokes parameters in Fourier space \cite{1997PhRvD..55.1830Z,2001PhRvD..64j3001Z,2016ARA&A..54..227K}
\begin{eqnarray}
    \Tilde E_{\nu}(\textbf{k}) =  \Tilde Q_{\nu}(\textbf{k})\cos{2\psi_{\textbf{k}}} + \Tilde U_{\nu}(\textbf{k})\sin{2\psi_{\textbf{k}}}
    , \nonumber \\
    \Tilde B_{\nu}(\textbf{k}) =  -\Tilde Q_{\nu}(\textbf{k})\sin{2\psi_{\textbf{k}}} + \Tilde U_{\nu}(\textbf{k})\cos{2\psi_{\textbf{k}}},
    \label{eq:E_B}
\end{eqnarray}
where $\psi_{\textbf{k}}$ is the angle between the wave-vector $\textbf{k}$ and line-of-sight.

We calculate the auto-angular power spectra for $I$, $E$, and $B$ radiance at the redshift bin centered at $z=8.0$ and $\Delta z = 0.5$ using the Limber approximation \cite{1953ApJ...117..134L} 
\begin{eqnarray}
    C_{\ell}^{I, E, B} &=& \int \frac{d\chi}{\chi^2} P_{I,E,B}(k=\frac{\ell+1/2}{\chi})
    \nonumber \\
     &=& \frac{\Delta z}{\chi^2}\frac{c}{H(z)}\left[\frac{H(z)\nu_{\rm obs}}{c(1+z)}\right]^2 P_{I_{\nu},E_{\nu},B_{\nu}}\left(k=\frac{\ell+1/2}{\chi}\right) ,
    \label{eq:Cl}
\end{eqnarray}
where $\chi(z=8)\approx 9136.2$ Mpc is the angular distance, the range of $\ell$ in this is $600\lesssim \ell\lesssim 2\times10^4$ with lower and upper limit determined by the simulation box length and cell size respectively, and we have used the formula for the radiance in a redshift bin
\begin{equation}
I =\int I_{\nu}\,d\nu=\int I_{\nu}\frac{H(z)\nu_{\rm obs}}{c(1+z)}\,d\chi
\end{equation}
in order to get the radial weight in the Limber integral (Eq.~\ref{eq:Cl}).

\subsection{Convergence Tests}
\label{ss:conv}

\begin{figure}
    \centering
    \includegraphics{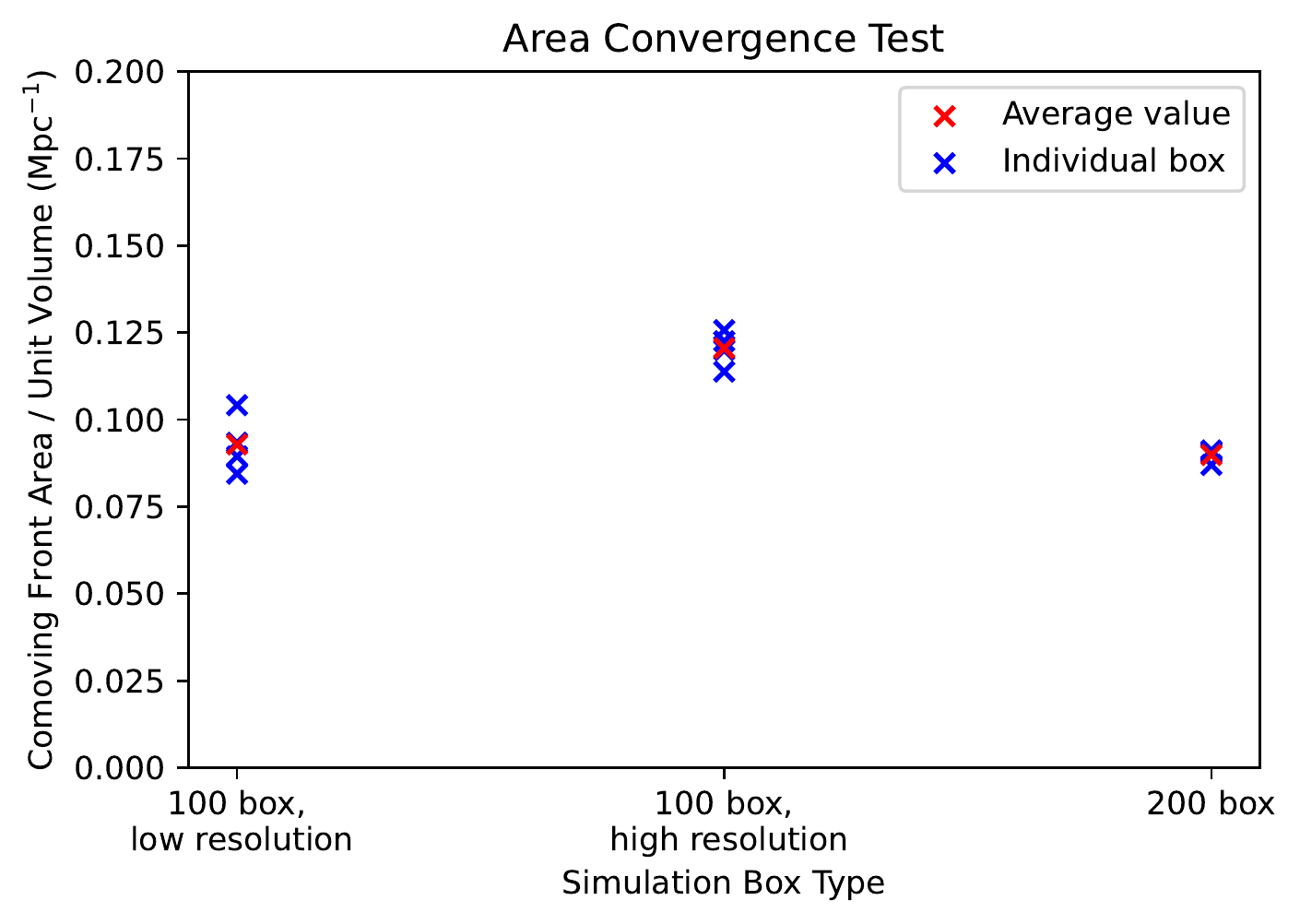}
    \caption{\large Convergence of comoving ionization front area per unit volume. Each simulation box is shown, as well as the average for the box types. to compare volumes of 100 Mpc$^{3}$.
    The higher resolution box has a slightly larger average than the low resolution box, as expected since more structure in the ionization fronts can be resolved, but as we are interested in an order of magnitude estimate of the power spectrum for detectability purposes, the areas are appropriately convergent for our purposes.}
    \label{fig:AreaConvergence}
\end{figure}

\begin{figure}
    \centering
    \subfloat[]{\includegraphics[width=73mm]{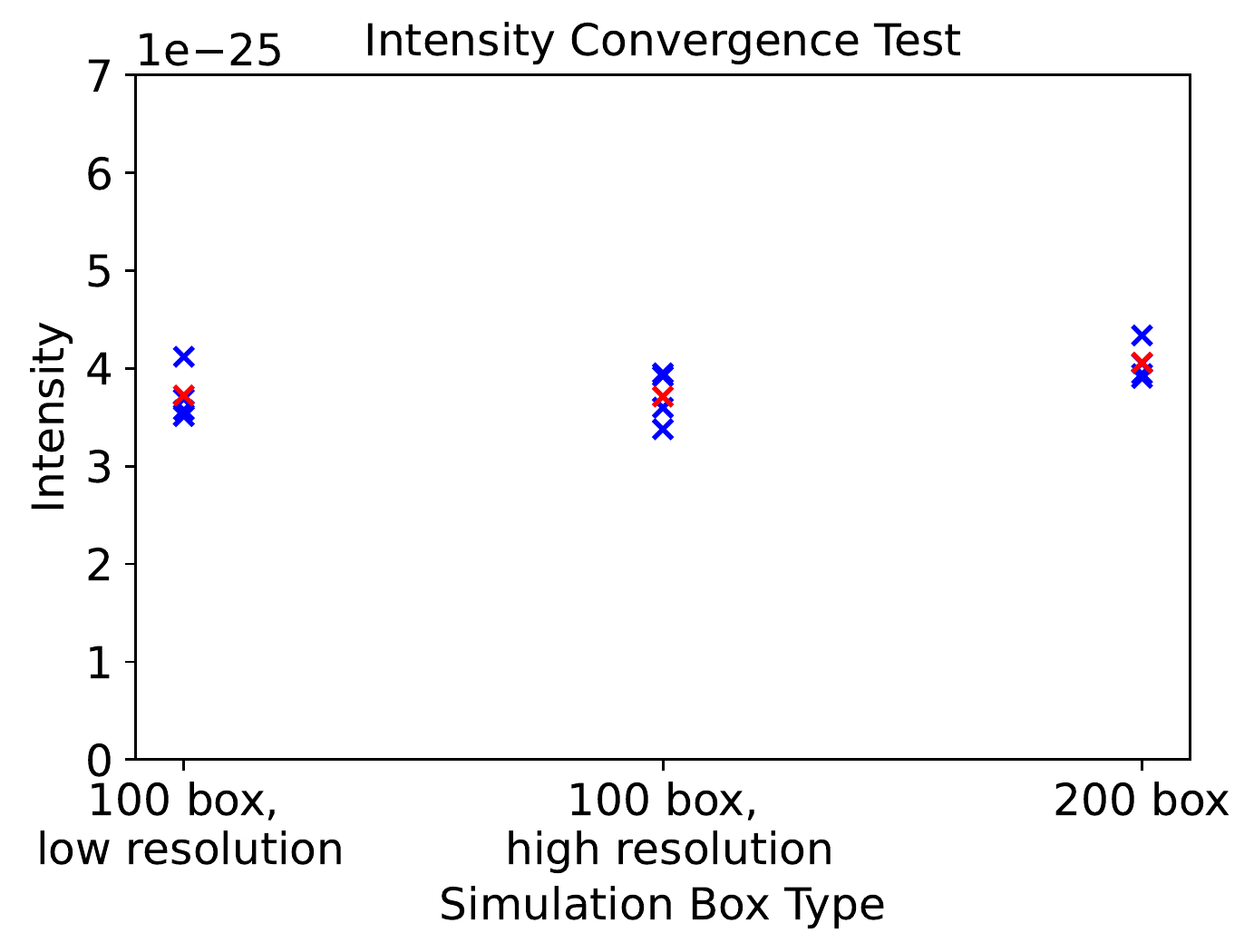}}
    \subfloat[]{\includegraphics[width=73mm]{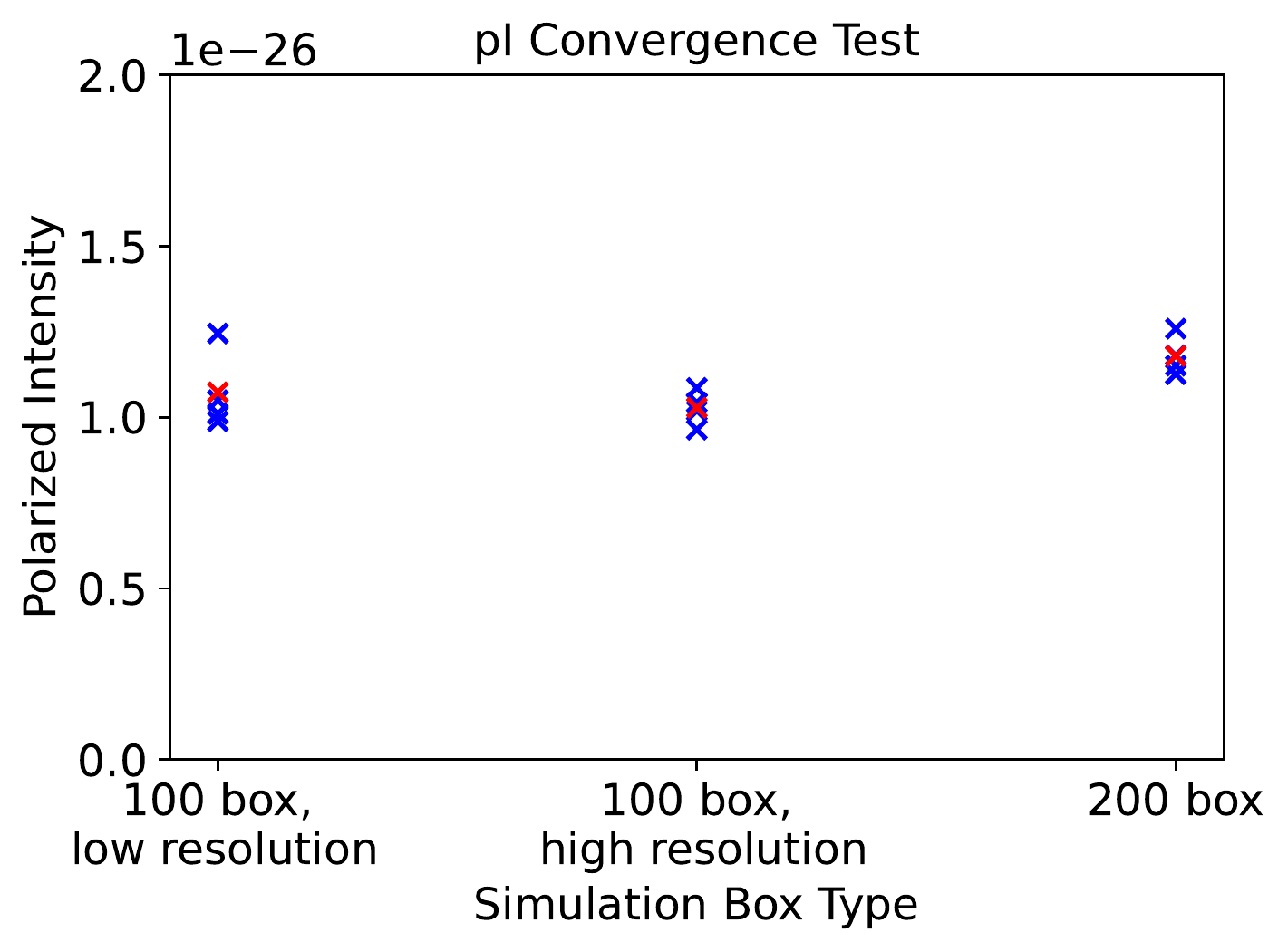}}
    \hspace{0mm}
    \subfloat[]{\includegraphics[width=77mm]{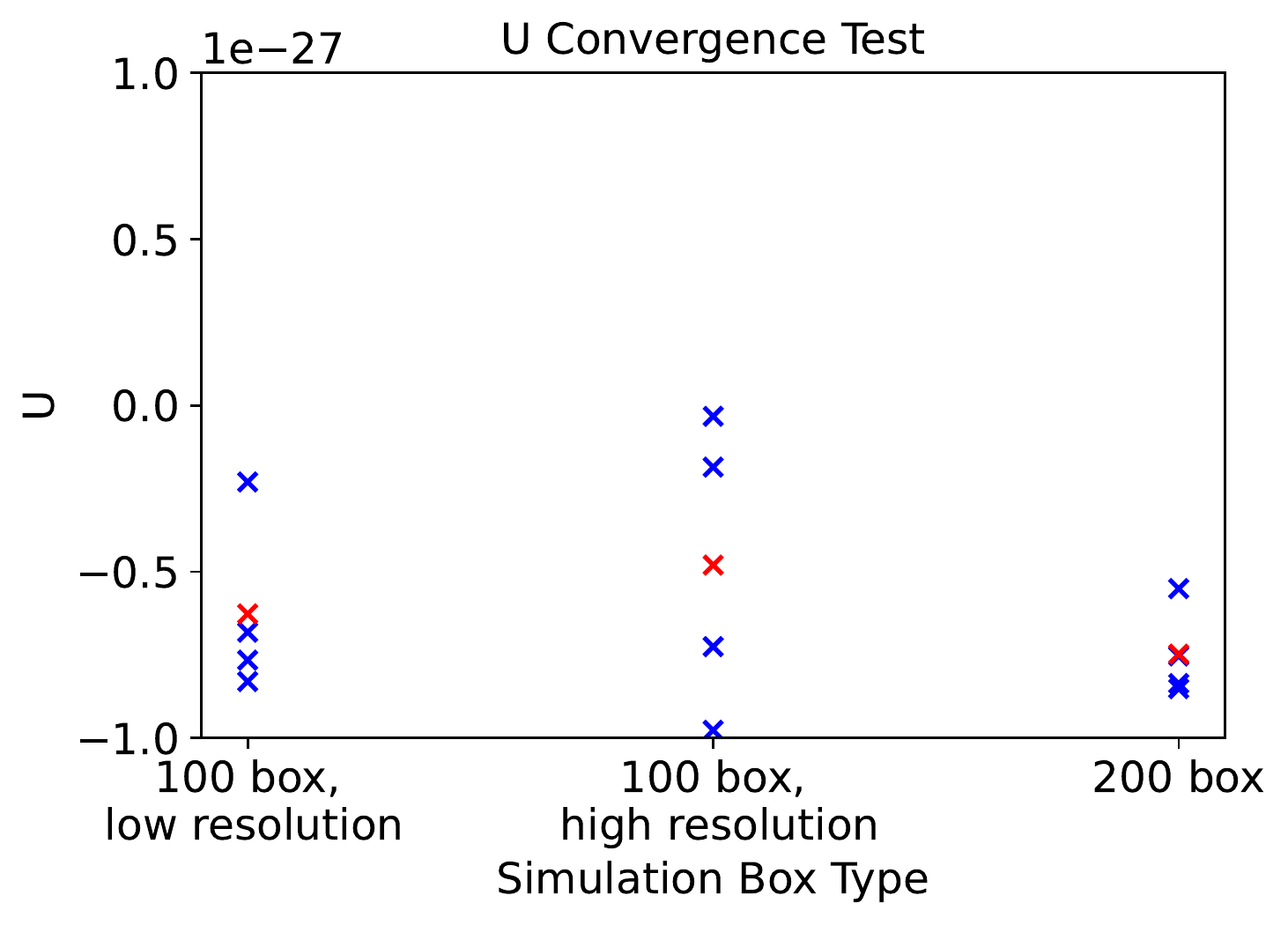}}
    \subfloat[]{\includegraphics[width=77mm]{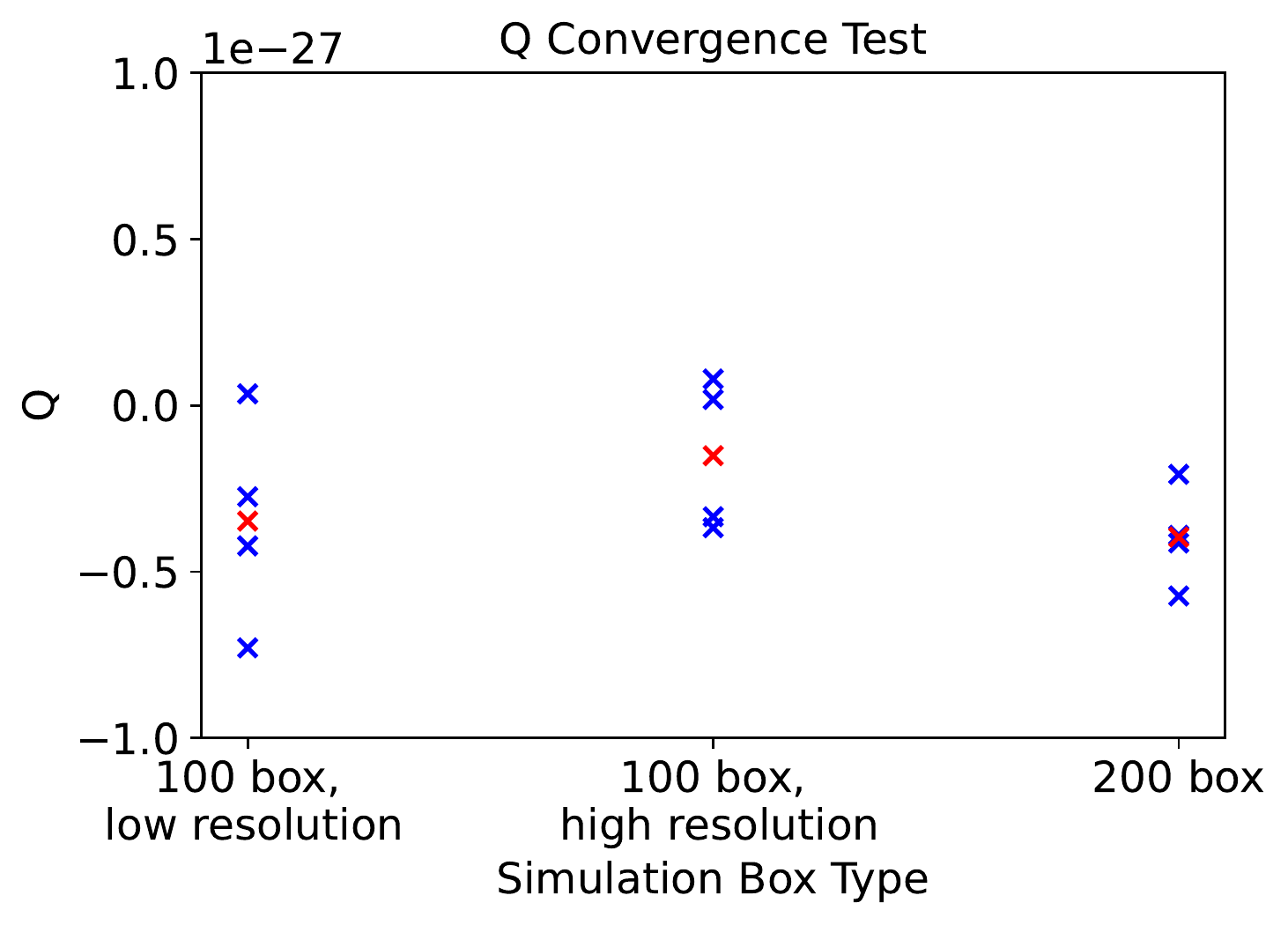}}
    \caption{Intensities convergence tests. Blue marks indicate values for each individual box, and red marks show the average value for each box type. Values shown in units  erg/cm$^{2}$/s/sr.}
    \label{fig:ConvergenceIntensities}
\end{figure}

As mentioned in section \ref{ss:21cmFAST}, we verify the intensity calculations by creating boxes with varying dimension and resolution. We generated 4 simulation boxes in each of the categories: (100 Mpc)$^3$ cube volume with (1 Mpc)$^3$ cell size, (100 Mpc)$^3$ cube volume with (0.5 Mpc)$^3$ cell size, and (200 Mpc)$^3$ cube volume with (1 Mpc)$^3$ cell size. For each of these categories, we compare the aggregate area of the fronts to ensure the convergence of the front modeling method. We again expect consistent results regardless of resolution size of the cells, and expect the area of the fronts to increase by a factor of 8 for the larger sized coeval box. The mean total area of the ionization fronts given by all four (100 Mpc)$^{3}$ low resolution boxes is 92900 Mpc$^{2}$, the higher resolution (100 Mpc)$^{3}$ boxes have an averaged aggregate area of 120000 Mpc$^{2}$, and the larger (200 Mpc)$^{3}$ box gives an averaged total front area of 719000 Mpc$^{2}$. Figure \ref{fig:AreaConvergence} shows the distribution of area and average area for each category of box. These values are within the expectations for the limits of this analysis so we consider this to be an acceptable convergence.

\begin{table}
    \begin{center}
    \begin{tabular}{|c|c|c|c|c|}
    \hline
    Box Type & I & pI & U & Q  \\
    \hline
   100 low res & 3.7233$\times$10$^{-25}$ & 1.0734$\times$10$^{-26}$& -6.2690 $\times$ 10$^{-28}$ &-3.4761$\times$10$^{-28}$ \\
    &$\pm$2.3723$\times$10$^{-26}$ & $\pm$1.0152$\times$10$^{-27}$ & $\pm$2.3511$\times$10$^{-28}$ & $\pm$2.7512$\times$10$^{-28}$\\
    \hline
   
   200 & 4.0588$\times$10$^{-25}$ & 1.1793$\times$10$^{-26}$&  -7.4780$\times$10$^{-28}$ &-3.9593$\times$10$^{-28}$ \\
    & $\pm$1.6981$\times$10$^{-26}$ & $\pm$4.9992$\times$10$^{-28}$ & $\pm$1.1985$\times$10$^{-28}$ & $\pm$1.2969$\times$10$^{-28}$\\
   \hline
   100 high res & 3.7118$\times$ 10$^{-25}$ & 1.0280$\times$ 10$^{-26}$&-4.7996$\times$ 10$^{-28}$&-1.5114$\times$10$^{-28}$\\ 
   & $\pm$2.3711$\times$10$^{-26}$ & $\pm$4.4062$\times$10$^{-28}$ & $\pm$3.8530$\times$10$^{-28}$ & $\pm$2.0138$\times$10$^{-28}$\\
   \hline
   \end{tabular}
   \caption{Intensity means and standard deviations for each box type in units  erg/cm$^{2}$/s/sr. Values are determined by computing the average over volume for each box independently, and taking the mean and standard deviation among the 4 boxes in the same category.}
   \label{intensity_chart}
   \end{center}
\end{table}

We also compare the mean intensity and the polarized intensity for each front. Consistent values for each type of simulation cube indicates a convergence in the intensity calculation based on size of simulation and resolution. Values for means and standard deviations of the different types of intensities for each box type can be found in table \ref{intensity_chart}. A figure demonstrating the spread of values for intensities for each box can be found in figure \ref{fig:ConvergenceIntensities}. We find the convergence of the intensity values is within the acceptable limit of our error, so we proceed with the calculation of the power spectrum. There is a slight asymmetry in the $\phi$ direction of the fronts with a preference for North-South and Northeast-Southwest orientations over Northwest-Southeast and East-West configurations of fronts, which is due to a directional preference for line segments of fronts in the 2D slabs of the simulation. As $\langle Q \rangle$ and $\langle U\rangle$ depend on $\cos(2\phi)$ and $\sin(2\phi)$, $\langle Q \rangle$ and $\langle U\rangle$ have a slight inconistency with 0. As the fractional inconsistency $\langle Q \rangle /\langle pI \rangle$ is less that 0.04 and $\langle U \rangle /\langle pI \rangle$ is less than .07, this is not a significant enough impact to effect our further data analysis. 

\section{Results}
\subsection{Angular power spectra}
\begin{figure}[h]
\centering
\includegraphics[width=6in]{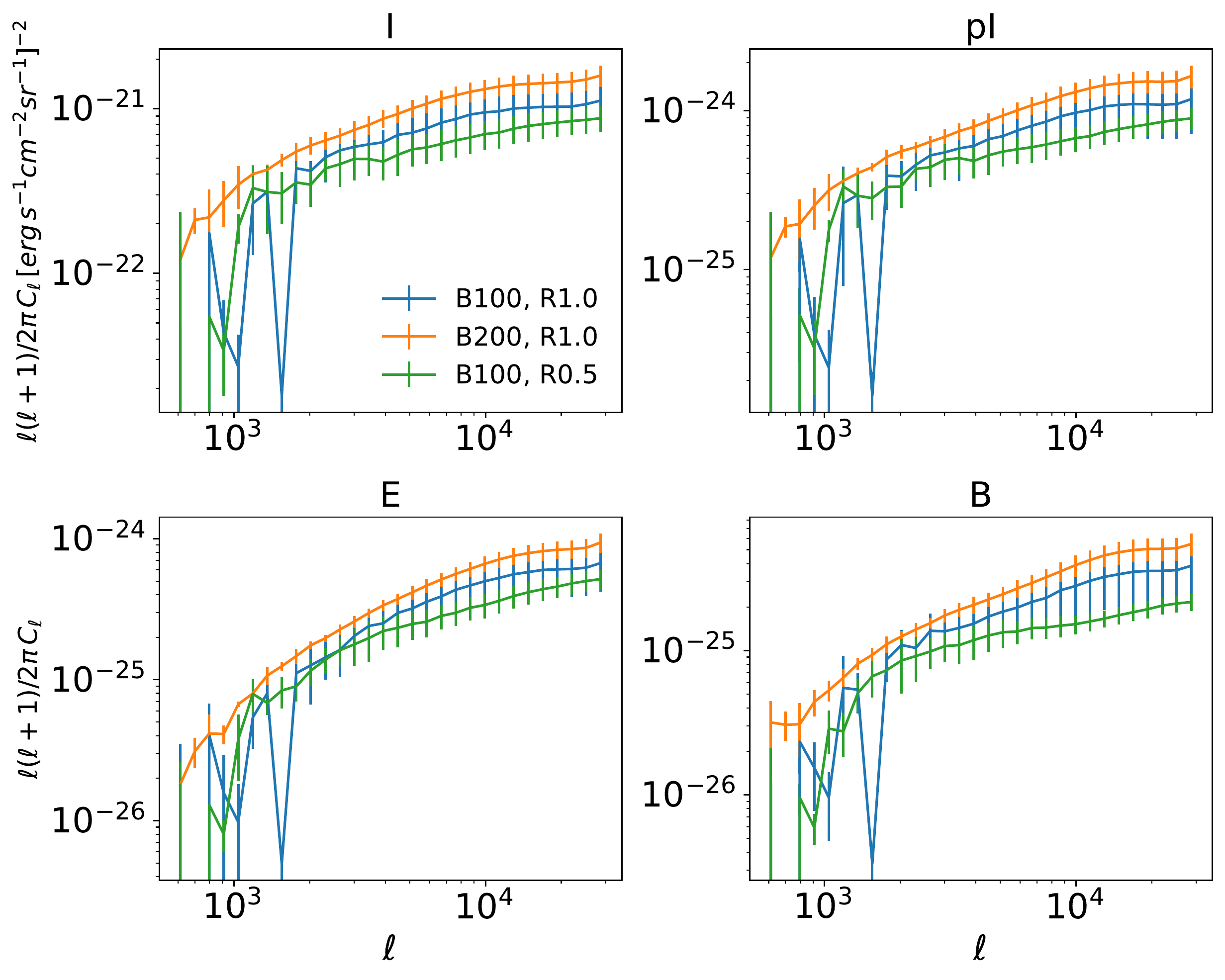}
\caption{Angular power spectra of \Lya\ total intensity I, polarized intensity pI, and that of E and B modes polarization from the cosmic ionization fronts at z=8.0 with a redshift depth $\Delta z=0.5$. Blue lines represents results from {\sc 21cmFAST} box with box length $L_{\rm box}=100 \rm Mpc$ and resolution for each grid $L_{\rm grid}=1 \rm Mpc$, orange lines represent $L_{\rm box}=200 \rm Mpc$, $L_{\rm grid}=1 \rm Mpc$, and green lines are from boxes $L_{\rm box}=100 \rm Mpc$, $L_{\rm grid}=0.5 \rm Mpc$. Error bars show Monte Carlo scatter from 4 realizations.} 
\label{fig:Cl}
\centering
\end{figure}

We show the angular power spectra of \Lya\ total intensity I, polarized intensity pI, and that of $E$ and $B$-mode polarization from the ionizing fronts at $z=8.0$ with a redshift depth $\Delta z=0.5$ in Figure~\ref{fig:Cl}.

The predicted polarization signal in Figure~\ref{fig:Cl} can be compared to the Lyman-$\alpha$ polarization signal expected from high-redshift galaxies, where the Lyman-$\alpha$ radiation scatters in the galaxies' haloes as it escapes. Mas-Ribas \& Chang \cite{MasRibasChang} estimate that the galaxy signal has a much bluer slope than that predicted here for ionization fronts, which is unsurprising for a signal from individual haloes. Their Fig.~3 shows an $E$-mode signal roughly equal to ours ($\sim 7\times 10^{-25}\, [{\rm erg}\,{\rm cm}^{-2}\,{\rm s}^{-1}\,{\rm sr}^{-1}]^{-2}$) at $\ell \approx 1.5\times 10^4$. The signal rapidly drops at lower $\ell$, reaching a negligible level of $10^{-26}\, [{\rm erg}\,{\rm cm}^{-2}\,{\rm s}^{-1}\,{\rm sr}^{-1}]^{-2}$ at $\ell\approx 7000$. Thus we conclude that at the large scales $\ell<1.5\times 10^4$, the Lyman-$\alpha$ signal from ionization fronts is likely to be stronger than the signal from individual galaxies.

We further note that our calculations predict a $B$-mode as well as an $E$-mode, with $C_\ell^{BB}/C_\ell^{EE} \approx 0.59 $ at $\ell\sim 10^4$. This result indicates that there is nonlinear scattering occurring in the ionization fronts, and also gives an indication of the geometry of the sources. The $B$-modes cannot be generated by a spherically symmetric object due to rotational symmetry requirements, but we do expect the sources to have deviations from spherical symmetry. The model for scattering in high-redshift galaxy haloes of Mas-Ribas \& Chang \cite{MasRibasChang} does not predict the $B$-mode because the haloes are taken to be spherically symmetric, and thus have the characteristic reflection symmetry across the Fourier wavevector that leads to $B=0$. Sources that are not spherically symmetric, but have polarization direction related to the geometry, have been studied in the CMB foreground literature and usually have both $E$ and $B$, but with $C_\ell^{BB}<C_\ell^{EE}$ \cite{2001PhRvD..64j3001Z}. We expect that inclusion of a realistic non-spherically symmetric model would produce a nonzero $B$-mode from galactic haloes, but probably would not change the basic conclusion that at large scales ($\ell\lesssim 10^4$) the total ($E+B$) Lyman-$\alpha$ polarization signal from the ionization fronts exceeds that from the galaxies.

\label{Sec:results}

\subsection{Detectability}

As there are no polarized Lyman-$\alpha$ intensity mapping surveys currently planned in the relevant wavelength range ($\lambda_{\rm obs} = \lambda_{{\rm Ly}\alpha}(1+z) \sim 1.1\,\mu$m), we consider the detectability limit for a few potential experiments. In general, we can consider two limits. For a noise-dominated experiment with a noise power spectrum $C_\ell^{\rm noise}$, covering a sky area $f_{\rm sky}$, the $Z\sigma$ detectability limit (one often considers $Z=3$ or 5) for auto-power in a bin of width $\Delta\ell$ is
\begin{equation}
C_\ell^{\rm detectable, auto} = \frac Z {\sqrt{f_{\rm sky}\ell\,\Delta\ell}} \,C_\ell^{\rm noise}.
\end{equation}
The $Z\sigma$ detectability limit in cross-correlation with a template that has a cross-correlation coefficient $\rho_\ell = C^{{\rm Ly}\alpha, X}_\ell / \sqrt{C^{{\rm Ly}\alpha, {\rm Ly}\alpha}_\ell C^{X X}_\ell}$ is
\begin{equation}
C_\ell^{\rm detectable, cross} = \frac{\rho_\ell^{-2} Z^2}{2f_{\rm sky}\ell\,\Delta\ell} \,C_\ell^{\rm noise}.
\end{equation}
The idealized ``perfect'' template, if external observations were able to provide full information on where to expect the ionization fronts, would have $\rho_\ell^2=1$. Even when such a template is not perfect, there may be both a signal-to-noise and a foreground rejection advantage to the cross-correlation method (particularly given that there are other potential polarized intensity mapping signals in this band \cite{MasRibasChang}).

We describe the noise model by comparison to the ``Lyapol-S'' experiment discussed in Mas-Ribas \& Chang \cite{MasRibasChang}. This is a concept based on the Cosmic Dawn Intensity Mapper (CDIM) proposal \cite{2019BAAS...51g..23C}. CDIM would use an effective 83 cm aperture telescope to feed a mosaic of infrared detectors at 1 arcsec plate scale covering 7.8 deg$^2$, using a linear variable filter to select a bandpass that varies over the focal plane. Stepping of the instrument along the sky allows a spectrum of each pixel to be built up. Lyapol-S would have to include a polarization capability as well; there are several ways to do this, but for the purposes of sensitivity calculations we will assume a linear polarizing filter and a half wave plate that rotates between exposures. The zodiacal sky brightness at the ecliptic poles is $I_\nu^{\rm sky} \approx 9.4\times 10^{-19}$ erg cm$^{-2}$ s$^{-1}$ sr$^{-1}$ Hz$^{-1}$ \cite{1998A&AS..127....1L}. On scales large compared to the beam size ($\ell<10^5$, considered here), the noise power spectrum per polarization component ($E$ or $B$) is
\begin{equation}
[C_\ell^{\rm noise}]^{1/2} =  \nu \sqrt{\frac{4f_{\rm n} hI_\nu^{\rm sky}}{ \eta A R t_0}},
\end{equation}
where $h$ is Planck's constant, $f_{\rm n}$ is the ratio of total noise variance (including, e.g., dark current and read noise) to sky noise, $\eta$ is the throughput, $A$ is the collecting area, $R$ is the spectral resolution, and $t_0$ is the observing time per spectral channel (assumed summed over 4 polarization angles: 0$^\circ$, 45$^\circ$, 90$^\circ$, and 135$^\circ$). If one could achieve $\eta/f_{\rm n} = 0.5$, i.e., 50\% effective throughput including degradation by other noise sources, and assumes the 83 cm aperture in analogy to CDIM, then the Lyapol-S sensitivity of $5.5\times 10^{-13}$\,erg cm$^{-2}$ s$^{-1}$ sr$^{-1/2}$ at $R=18$ could be achieved with $t_0 = 1.26\times 10^5\,$s. Even in a highly purpose-optimized experiment with, e.g., 20 bands across the near infrared, a 300 deg$^2$ survey as proposed for Lyapol-S would require a total live time of $(300/7.8)\times 20 \times( 1.26\times 10^5\,{\rm s})$, or 3 years. We therefore regard Lyapol-S as a particularly ambitious survey concept, although one where the raw sensitivity appears to be achievable with present technology.

Proceeding with Lyapol-S, and scaled to a bin width of $\Delta\ell/\ell = 0.5$ and detection significance $Z=3$, the detectability thresholds are
\begin{equation}
\frac{\ell(\ell+1)}{2\pi}C_\ell^{\rm detectable, auto} = 2.3\times 10^{-21} \left( \frac{Z}{3} \right) \left( \frac{0.5}{\Delta \ell/\ell} \right)^{1/2} \left(\frac{\ell}{1000} \right)\,{\rm erg\,cm^{-2}\,s^{-1}\,sr^{-1}}
\end{equation}
and
\begin{equation}
\frac{\ell(\ell+1)}{2\pi}C_\ell^{\rm detectable, cross} = 5.4\times 10^{-23} \rho_\ell^{-2} \left( \frac{Z}{3} \right)^2 \left( \frac{0.5}{\Delta \ell/\ell} \right)
\,{\rm erg\,cm^{-2}\,s^{-1}\,sr^{-1}}.
\label{eq:C-cross}
\end{equation}

We can see that even a detection in cross correlation with an (obviously idealized) perfect template, $|\rho_\ell|=1$, requires a factor of $\sim 100$ improvement in noise power $C_\ell^{\rm noise}$ relative to Lyapol-S. It therefore appears that some relatively major advances -- e.g., development of a energy-resolving detector in the NIR that would enable all of the bands to be measured at once everywhere in the focal plane (in principle a factor of $\sim 20$, although the energy resolution in the 1 $\mu$m band would need to be improved beyond the current generation of photon-counting detectors \cite{2019BAAS...51g..17M}) and going to the outer solar system or out of the Ecliptic Plane where the sky background is lower (where an order of magnitude or more is possible \cite{1979A&A....77..223G}). Therefore, while there are no fundamental physical principles that would prevent us from reaching the required sensitivity with a $\sim 1$ m class space telescope, we see it as a rather futuristic concept.

\section{Discussion}
\label{sec:Discussion}

We have made a first estimate of \Lya\ polarization arising from the cosmic ionization fronts at $z=8$ in this work. We have used the microscopic physics model and Monte-Carlo simulations of \Lya\ photons passing through ionization fronts in Paper I and developed a methodology to extract the ionization fronts from {\sc 21cmFAST} simulations. We have estimated the auto power spectra of the \Lya\ quantities total intenstiy, I, polarized intensity, pI, and the \Lya\ E and B modes, from the cosmic ionization fronts at redshift 8. In order to assess detectablity scales, we compare our results with proposed ``Lyapol-S'' experiment proposed in Mas-Ribas \& Chang \cite{MasRibasChang}, specifically investigating the expected auto-power and cross-correlations from their proposal. From this, we find that even with a highly specialized survey proposal we would need significant advances in order to map the ionizing fronts by \Lya\ polarization intensity mapping.

While we do not anticipate a survey for polarized \Lya\ emission from reionization to be completed in the near future, we believe that technology will eventually advance enough to accomplish this goal, and therefore the further investigation of this potential signal is justified. For future work, one could explore the tomography of \Lya\ polarization by extending the estimation at $z=8$ in this work to the whole Epoch of Reionization. The method of extracting ionization fronts could be implemented to calculate other signals arising from cosmic ionization fronts. Furthermore, a more detailed investigation of the cross-correlation of \Lya\ polarization signal with other tracers sensitive to the neutral hydrogen fraction during the reionization (e.g., 21 cm) could help us understand the realistic range of correlation coefficients $\rho_\ell$ 
at the scales of ionization bubbles and thus the detectability of the ionization fronts in cross-correlation.

\section*{Acknowledgements}

We thank Tzu-Ching Chang and Chenxiao Zeng for useful feedback on the draft of this paper.

During the preparation of this work, the authors were supported by NASA award 15-WFIRST15-0008, Simons Foundation award 60052667, and the David \& Lucile Packard Foundation.

This article used resources on the Pitzer Cluster at the Ohio Supercomputing Center \cite{OSC}.

\section*{Data Availability}
The code and data supporting this article may be made available on reasonable request to the corresponding author. 

\bibliographystyle{JHEP.bst}
\bibliography{main.bib}

\providecommand{\href}[2]{#2}\begingroup\raggedright\begin{thebibliography}{10}

\bibitem{2013ApJ...763..132S}
M.~B. {Silva}, M.~G. {Santos}, Y.~{Gong}, A.~{Cooray} and J.~{Bock},
  \emph{{Intensity Mapping of Ly{\ensuremath{\alpha}} Emission during the Epoch
  of Reionization}},
  \href{https://doi.org/10.1088/0004-637X/763/2/132}{\emph{\apj} {\bfseries
  763} (Feb., 2013) 132}, [\href{https://arxiv.org/abs/1205.1493}{{\ttfamily
  1205.1493}}].

\bibitem{Pullen_2014}
A.~R. Pullen, O.~Dor{\'{e} } and J.~Bock, \emph{{Intensity mapping across
  cosmic times with the Ly$\alpha$ line}},
  \href{https://doi.org/10.1088/0004-637x/786/2/111}{\emph{The Astrophysical
  Journal} {\bfseries 786} (apr, 2014) 111}.

\bibitem{2017ApJ...848...52H}
C.~{Heneka}, A.~{Cooray} and C.~{Feng}, \emph{{Probing the Intergalactic Medium
  with Ly{\ensuremath{\alpha}} and 21 cm Fluctuations}},
  \href{https://doi.org/10.3847/1538-4357/aa8eed}{\emph{\apj} {\bfseries 848}
  (Oct., 2017) 52}, [\href{https://arxiv.org/abs/1611.09682}{{\ttfamily
  1611.09682}}].

\bibitem{2022arXiv221009612S}
C.~{Shekhar Murmu}, R.~{Ghara}, S.~{Majumdar} and K.~K. {Datta}, \emph{{Probing
  the Epoch of Reionization using synergies of line intensity mapping}},
  \href{https://doi.org/10.48550/arXiv.2210.09612}{\emph{arXiv e-prints} (Oct.,
  2022) arXiv:2210.09612}, [\href{https://arxiv.org/abs/2210.09612}{{\ttfamily
  2210.09612}}].

\bibitem{PaperI}
Y.~Yang, E.~Koivu, C.~Zeng, H.~Long and C.~M. Hirata, \emph{Lyman-$\alpha$
  polarization from cosmological ionization fronts: I. radiative transfer
  simulations}, {\emph{in prep.} (feb, 2023) }.

\bibitem{https://doi.org/10.48550/arxiv.1709.09066}
E.~D. Kovetz, M.~P. Viero, A.~Lidz, L.~Newburgh, M.~Rahman, E.~Switzer et~al.,
  \emph{Line-intensity mapping: 2017 status report},  2017.
\newblock 10.48550/ARXIV.1709.09066.

\bibitem{2008ASPC..399..115H}
G.~J. {Hill}, K.~{Gebhardt}, E.~{Komatsu}, N.~{Drory}, P.~J. {MacQueen},
  J.~{Adams} et~al., \emph{{The Hobby-Eberly Telescope Dark Energy Experiment
  (HETDEX): Description and Early Pilot Survey Results}},  in \emph{Panoramic
  Views of Galaxy Formation and Evolution} (T.~{Kodama}, T.~{Yamada} and
  K.~{Aoki}, eds.), vol.~399 of \emph{Astronomical Society of the Pacific
  Conference Series}, p.~115, Oct., 2008,
  \href{https://arxiv.org/abs/0806.0183}{{\ttfamily 0806.0183}},
  \href{https://doi.org/10.48550/arXiv.0806.0183}{DOI}.

\bibitem{2021ApJ...923..217G}
K.~{Gebhardt}, E.~{Mentuch Cooper}, R.~{Ciardullo}, V.~{Acquaviva},
  R.~{Bender}, W.~P. {Bowman} et~al., \emph{{The Hobby-Eberly Telescope Dark
  Energy Experiment (HETDEX) Survey Design, Reductions, and Detections}},
  \href{https://doi.org/10.3847/1538-4357/ac2e03}{\emph{\apj} {\bfseries 923}
  (Dec., 2021) 217}, [\href{https://arxiv.org/abs/2110.04298}{{\ttfamily
  2110.04298}}].

\bibitem{2021MNRAS.501.3883R}
P.~{Renard}, E.~{Gaztanaga}, R.~{Croft}, L.~{Cabayol}, J.~{Carretero},
  M.~{Eriksen} et~al., \emph{{The PAU survey: Ly {\ensuremath{\alpha}}
  intensity mapping forecast}},
  \href{https://doi.org/10.1093/mnras/staa3783}{\emph{\mnras} {\bfseries 501}
  (Mar., 2021) 3883--3899}, [\href{https://arxiv.org/abs/2006.07177}{{\ttfamily
  2006.07177}}].

\bibitem{2014arXiv1412.4872D}
O.~{Dor{\'e}}, J.~{Bock}, M.~{Ashby}, P.~{Capak}, A.~{Cooray}, R.~{de Putter}
  et~al., \emph{{Cosmology with the SPHEREX All-Sky Spectral Survey}},
  \href{https://doi.org/10.48550/arXiv.1412.4872}{\emph{arXiv e-prints} (Dec.,
  2014) arXiv:1412.4872}, [\href{https://arxiv.org/abs/1412.4872}{{\ttfamily
  1412.4872}}].

\bibitem{Loeb_1999}
A.~Loeb and G.~B. Rybicki, \emph{Scattered ly$\alpha$ radiation around sources
  before cosmological reionization},
  \href{https://doi.org/10.1086/307844}{\emph{The Astrophysical Journal}
  {\bfseries 524} (oct, 1999) 527--535}.

\bibitem{Dijkstra_2008}
M.~Dijkstra and A.~Loeb, \emph{The polarization of scattered ly{\^{i} }$\pm$
  radiation around high-redshift galaxies},
  \href{https://doi.org/10.1111/j.1365-2966.2008.13066.x}{\emph{Monthly Notices
  of the Royal Astronomical Society} {\bfseries 386} (may, 2008) 492--504}.

\bibitem{MasRibasChang}
L.~{Mas-Ribas} and T.-C. {Chang}, \emph{Lyman-$\ensuremath{\alpha}$
  polarization intensity mapping},
  \href{https://doi.org/10.1103/PhysRevD.101.083032}{\emph{Phys. Rev. D}
  {\bfseries 101} (Apr, 2020) 083032}.

\bibitem{Murray2020}
S.~G. Murray, B.~Greig, A.~Mesinger, J.~B. Muñoz, Y.~Qin, J.~Park et~al.,
  \emph{21cmfast v3: A python-integrated c code for generating 3d realizations
  of the cosmic 21cm signal.},
  \href{https://doi.org/10.21105/joss.02582}{\emph{Journal of Open Source
  Software} {\bfseries 5} (2020) 2582}.

\bibitem{10.1111/j.1365-2966.2010.17731.x}
A.~Mesinger, S.~Furlanetto and R.~Cen, \emph{{21cmfast: a fast, seminumerical
  simulation of the high-redshift 21-cm signal}},
  \href{https://doi.org/10.1111/j.1365-2966.2010.17731.x}{\emph{Monthly Notices
  of the Royal Astronomical Society} {\bfseries 411} (02, 2011) 955--972},
  [\href{https://arxiv.org/abs/https://academic.oup.com/mnras/article-pdf/411/2/955/4099991/mnras0411-0955.pdf}{{\ttfamily
  https://academic.oup.com/mnras/article-pdf/411/2/955/4099991/mnras0411-0955.pdf}}].

\bibitem{2020A&A...641A...6P}
{Planck Collaboration}, N.~{Aghanim}, Y.~{Akrami}, M.~{Ashdown}, J.~{Aumont},
  C.~{Baccigalupi} et~al., \emph{{Planck 2018 results. VI. Cosmological
  parameters}}, \href{https://doi.org/10.1051/0004-6361/201833910}{\emph{\aap}
  {\bfseries 641} (Sept., 2020) A6},
  [\href{https://arxiv.org/abs/1807.06209}{{\ttfamily 1807.06209}}].

\bibitem{2017MNRAS.465.4838G}
B.~{Greig} and A.~{Mesinger}, \emph{{The global history of reionization}},
  \href{https://doi.org/10.1093/mnras/stw3026}{\emph{\mnras} {\bfseries 465}
  (Mar., 2017) 4838--4852}, [\href{https://arxiv.org/abs/1605.05374}{{\ttfamily
  1605.05374}}].

\bibitem{2018ApJ...864..142D}
F.~B. {Davies}, J.~F. {Hennawi}, E.~{Ba{\~n}ados}, Z.~{Luki{\'c}},
  R.~{Decarli}, X.~{Fan} et~al., \emph{{Quantitative Constraints on the
  Reionization History from the IGM Damping Wing Signature in Two Quasars at z
  > 7}}, \href{https://doi.org/10.3847/1538-4357/aad6dc}{\emph{\apj} {\bfseries
  864} (Sept., 2018) 142}, [\href{https://arxiv.org/abs/1802.06066}{{\ttfamily
  1802.06066}}].

\bibitem{Greig_2015}
B.~Greig and A.~Mesinger, \emph{21cmmc: an {MCMC} analysis tool enabling
  astrophysical parameter studies of the cosmic 21~cm signal},
  \href{https://doi.org/10.1093/mnras/stv571}{\emph{Monthly Notices of the
  Royal Astronomical Society} {\bfseries 449} (apr, 2015) 4246--4263}.

\bibitem{1999ApJS..123....3L}
C.~{Leitherer}, D.~{Schaerer}, J.~D. {Goldader}, R.~M.~G. {Delgado},
  C.~{Robert}, D.~F. {Kune} et~al., \emph{{Starburst99: Synthesis Models for
  Galaxies with Active Star Formation}},
  \href{https://doi.org/10.1086/313233}{\emph{\apjs} {\bfseries 123} (July,
  1999) 3--40}, [\href{https://arxiv.org/abs/astro-ph/9902334}{{\ttfamily
  astro-ph/9902334}}].

\bibitem{2001MNRAS.322..231K}
P.~{Kroupa}, \emph{{On the variation of the initial mass function}},
  \href{https://doi.org/10.1046/j.1365-8711.2001.04022.x}{\emph{\mnras}
  {\bfseries 322} (Apr., 2001) 231--246},
  [\href{https://arxiv.org/abs/astro-ph/0009005}{{\ttfamily
  astro-ph/0009005}}].

\bibitem{1996ApJ...461...20H}
F.~{Haardt} and P.~{Madau}, \emph{{Radiative Transfer in a Clumpy Universe. II.
  The Ultraviolet Extragalactic Background}},
  \href{https://doi.org/10.1086/177035}{\emph{\apj} {\bfseries 461} (Apr.,
  1996) 20}, [\href{https://arxiv.org/abs/astro-ph/9509093}{{\ttfamily
  astro-ph/9509093}}].

\bibitem{2006agna.book.....O}
D.~E. {Osterbrock} and G.~J. {Ferland}, \emph{{Astrophysics of gaseous nebulae
  and active galactic nuclei}}.
\newblock 2006.

\bibitem{1972hmfw.book.....A}
M.~{Abramowitz} and I.~A. {Stegun}, \emph{{Handbook of Mathematical
  Functions}}.
\newblock 1972.

\bibitem{1997PhRvD..55.1830Z}
M.~{Zaldarriaga} and U.~{Seljak}, \emph{{All-sky analysis of polarization in
  the microwave background}},
  \href{https://doi.org/10.1103/PhysRevD.55.1830}{\emph{\prd} {\bfseries 55}
  (Feb., 1997) 1830--1840},
  [\href{https://arxiv.org/abs/astro-ph/9609170}{{\ttfamily
  astro-ph/9609170}}].

\bibitem{2001PhRvD..64j3001Z}
M.~{Zaldarriaga}, \emph{{Nature of the E-B decomposition of CMB polarization}},
  \href{https://doi.org/10.1103/PhysRevD.64.103001}{\emph{\prd} {\bfseries 64}
  (Nov., 2001) 103001},
  [\href{https://arxiv.org/abs/astro-ph/0106174}{{\ttfamily
  astro-ph/0106174}}].

\bibitem{2016ARA&A..54..227K}
M.~{Kamionkowski} and E.~D. {Kovetz}, \emph{{The Quest for B Modes from
  Inflationary Gravitational Waves}},
  \href{https://doi.org/10.1146/annurev-astro-081915-023433}{\emph{\araa}
  {\bfseries 54} (Sept., 2016) 227--269},
  [\href{https://arxiv.org/abs/1510.06042}{{\ttfamily 1510.06042}}].

\bibitem{1953ApJ...117..134L}
D.~N. {Limber}, \emph{{The Analysis of Counts of the Extragalactic Nebulae in
  Terms of a Fluctuating Density Field.}},
  \href{https://doi.org/10.1086/145672}{\emph{\apj} {\bfseries 117} (Jan.,
  1953) 134}.

\bibitem{2019BAAS...51g..23C}
A.~{Cooray}, T.-C. {Chang}, S.~{Unwin}, M.~{Zemcov}, A.~{Coffey},
  P.~{Morrissey} et~al., \emph{{Cosmic Dawn Intensity Mapper}},  in
  \emph{Bulletin of the American Astronomical Society}, vol.~51, p.~23, Sept.,
  2019, \href{https://arxiv.org/abs/1903.03144}{{\ttfamily 1903.03144}}.

\bibitem{1998A&AS..127....1L}
C.~{Leinert}, S.~{Bowyer}, L.~K. {Haikala}, M.~S. {Hanner}, M.~G. {Hauser},
  A.~C. {Levasseur-Regourd} et~al., \emph{{The 1997 reference of diffuse night
  sky brightness}}, \href{https://doi.org/10.1051/aas:1998105}{\emph{\aaps}
  {\bfseries 127} (Jan., 1998) 1--99}.

\bibitem{2019BAAS...51g..17M}
B.~{Mazin}, J.~{Bailey}, J.~{Bartlett}, C.~{Bockstiegel}, B.~{Bumble},
  G.~{Coiffard} et~al., \emph{{MKIDs in the 2020s}},  in \emph{Bulletin of the
  American Astronomical Society}, vol.~51, p.~17, Sept., 2019,
  \href{https://arxiv.org/abs/1908.02775}{{\ttfamily 1908.02775}}.

\bibitem{1979A&A....77..223G}
R.~H. {Giese}, \emph{{Zodiacal light and local interstellar dust: predictions
  for an out-of-ecliptic spacecraft.}}, {\emph{\aap} {\bfseries 77} (Aug.,
  1979) 223--226}.

\bibitem{OSC}
{Ohio Supercomputer Center}, \emph{Ohio supercomputer center},  1987.

\end{thebibliography}\endgroup

\appendix

\section{Mapping ionization front triangles onto a rectangular grid}
\label{appendix:split}

\begin{figure}
\centering
\subfloat[]{\includegraphics[height=60mm]{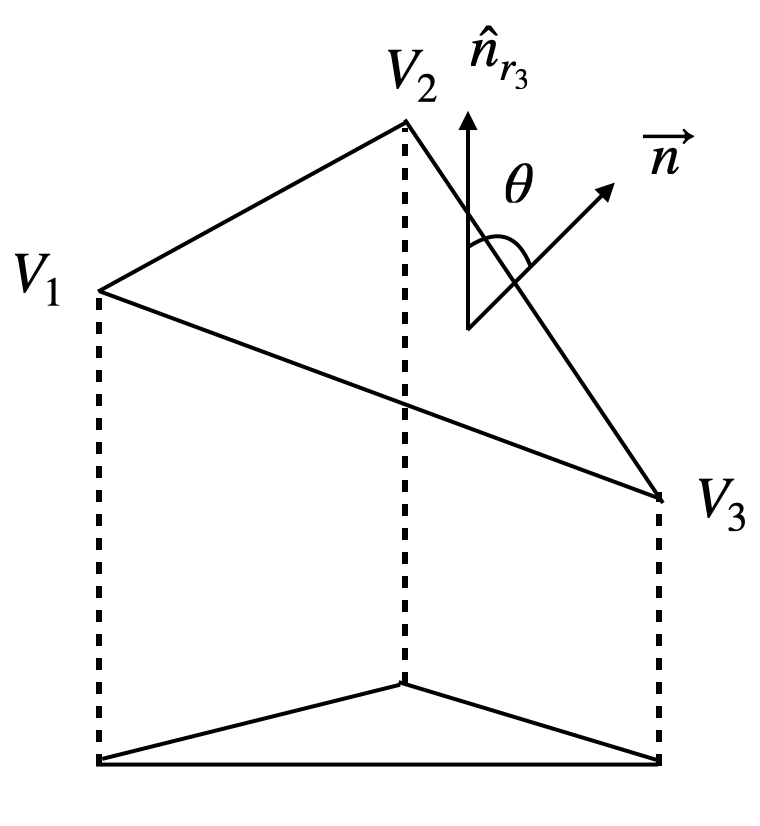}}
\quad
\subfloat[]{\includegraphics[height=60mm]{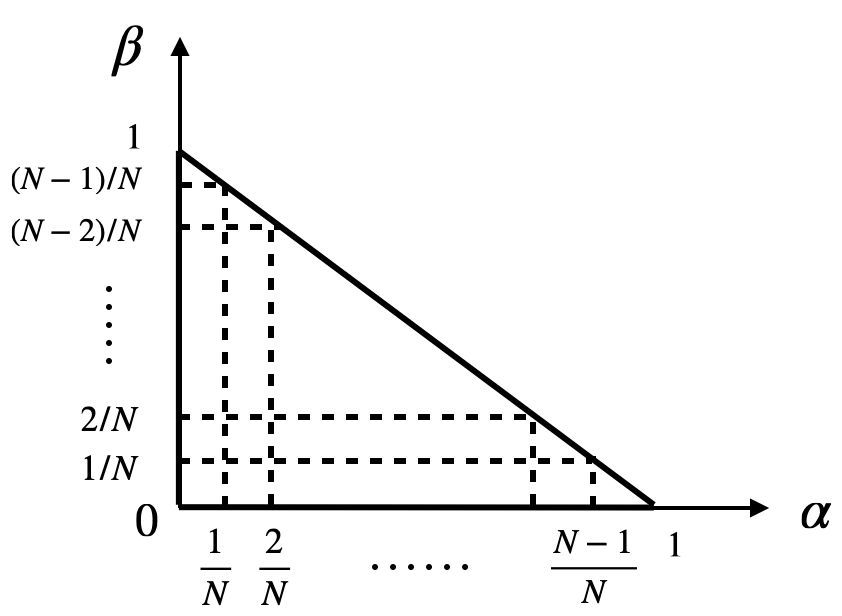}}
\caption{ (a) Projection of a triangle with vertices on $\Vec{V}_1,\,\Vec{V}_2,\,\Vec{V}_3$ to the $r_1-r_2$ plane such that a point $\Vec{P}$ on the triangle can be denoted by its projected 2D coordinate. (b) Parametrization of two sides coordinates by $\alpha$ and $\beta$.}
\centering
\label{fig:split_triangle}
\end{figure}

As demonstrated in Figure~\ref{fig:split_triangle}, in the $r_1,r_2,r_3$- Cartesian coordinate system, our method of splitting fronts back to grids involves parametrizing points on the triangle by two parameters $\alpha$ and $\beta$ along two sides of triangle projected to $r_1-r_2$ plane such that the point $\Vec{P}=\Vec{V}_1+\alpha(\Vec{V}_2-\Vec{V}_1)+\beta(\Vec{V}_3-\Vec{V}_1)$, and then add up the intensities from the split piece centered at the point to the grid containing this point.
Consider the specific intensity $I_{\nu,\Delta}$ and polarized specific intensity $Q_{\nu,\Delta}$, $U_{\nu,\Delta}$ of the triangle $\Delta$, the Eq. (4.4) and (4.5) in Paper I become

\begin{eqnarray}
    I_{\nu_{\rm obs},\Delta} &=& \frac{h\nu_{\rm obs} nP(\mu)}{2\pi(1+z)^3}\frac{(1+z)c}{H(z)\nu_{\rm obs}}\frac{\delta(r_3-r_{3,\rm front})}{V_{\rm grid}|\cos{\theta}|}\,\chi_{\Delta}(r_1,r_2)
    \nonumber \\
    Q_{\nu_{\rm obs},\Delta} &=& \frac{h\nu_{\rm obs} n\,pP(\mu)}{2\pi(1+z)^3}\frac{(1+z)c}{H(z)\nu_{\rm obs}}\frac{\delta(r_3-r_{3,\rm front})}{V_{\rm grid}|\cos{\theta}|}\chi_{\Delta}(r_1,r_2)\cos{2\phi}
     \\
    U_{\nu_{\rm obs},\Delta} &=& \frac{h\nu_{\rm obs} n\,pP(\mu)}{2\pi(1+z)^3}\frac{(1+z)c}{H(z)\nu_{\rm obs}}\frac{\delta(r_3-r_{3,\rm front})}{V_{\rm grid}|\cos{\theta}|}\chi_{\Delta}(r_1,r_2)\sin{2\phi}
    \nonumber
\end{eqnarray}
where $V_{\rm grid}$ is the volume for each gird, $\theta$ and $\phi$ is the polar and azimuthal angle of the point respectively. The characteristic function $\chi_{\Delta}$ loops over points on one triangle:
\begin{equation}
    \chi_{\Delta}(r_1,r_2)=2A_{\rm proj\,\Delta}\int^1_0d\alpha\int^{1-\alpha}_0 d\beta\,\delta(r_{1}-r_1(\alpha,\beta))\delta(r_{2}-r_2(\alpha,\beta))
\end{equation}
where $A_{\rm proj\Delta}=A_{\Delta}\,|\cos{\theta}|$ is the projected triangle area. Then for a grid at ($r_1,r_2,r_3$) we loop over all front triangles to count the specific intensity and so for polarized intensities
\begin{eqnarray}
    I_{\nu}(r_1,r_2,r_3) &=& \frac{h\nu_{\rm obs} n\,pP(\mu)}{2\pi(1+z)^3}\frac{(1+z)c}{H(z)\nu_{\rm obs}}\frac{1}{V_{\rm grid}}\sum_{\rm triangles}2A_{\Delta}\times \nonumber \\
    && \sum_{\alpha,\beta}\frac{1}{N^2}\times
    \begin{cases}
        1, \text{if point $\Vec{P}\in$ grid($r_1,r_2,r_3$) and is inside of $\Delta$} \\
        \frac{1}{2}, \text{if point $\Vec{P}\in$ grid($r_1,r_2,r_3$) and is on the sides of $\Delta$}\\
        \frac{1}{6}, \text{if point $\Vec{P}\in$ grid($r_1,r_2,r_3$) and is the vertex of $\Delta$}\\
        0, \text{if point $\Vec{P}\notin$ grid($r_1,r_2,r_3$)}
    \end{cases}
\end{eqnarray}
where the points contribute to the grid intensities fractionally depending on their positions on the sides, vertex or inside of the triangle $\Delta$ regarding points on the sides or vertex are shared by different triangles. We determine the value of N for each triangle by requiring the spaced interval for the triangle is smaller than the shortest wave mode with $k_{\rm max}=2\,\rm Mpc^{-1}$:
\begin{equation}
    \frac{L}{N}<\frac{\pi}{k_{\rm max}},
\end{equation}
where $L$ is the longest side of the triangle.

\end{document}